# La scala dell'energia

Anna Maria Aloisi

IPSIA "A. Meucci", Cagliari,

http:apmf.interfree.it, ampf@interfree.it, aloisiannamaria@progettomarte.net

Pier Franco Nali

*Dirigente con incarico di studio, ricerca e consulenza presso la Regione Sardegna, Cagliari,*

pnali@regione.sardegna.it

LE famose *Lezioni di Fisica* di Feynman si aprono con l'affermazione che l'ipotesi atomica, cioè l'idea che tutte le cose sono formate di atomi - minuscole particelle che si urtano e si respingono agitandosi in continuazione - rappresenta il concetto più importante, potente ed unificante di tutta la scienza. Come ci fa notare Feynman, si tratta di un'idea scientifica che si può esprimere con un piccolo numero di parole, ma comunica molte informazioni importanti sul mondo fisico; inoltre ci trasmette un'immagine intuitiva e al tempo stesso concreta e dinamica del mondo degli atomi, dandocene una prospettiva dall'alto che ricorda le straordinarie visioni cosmiche di Lucrezio e Democrito.

Ma se cambiamo prospettiva e consideriamo ciò che avviene *all'interno* di un atomo, qual è l'idea più importante che ci aiuta a capire il mondo a questo livello più profondo? È, secondo noi, l'ipotesi di Bohr che *un atomo può trovarsi soltanto in un numero determinato di stati permessi che obbedisce a condizioni specifiche - e non in uno qualsiasi degli infiniti stati possibili che ci dà la fisica classica - e compie transizioni soltanto fra questi stati discreti*. Ci è stato insegnato che gli stati atomici sono occupati da elettroni e che durante le transizioni gli elettroni *"saltano"* da uno stato a un altro, ma per il momento possiamo prescindere da questo dato e parlare di stati e di transizioni per l'atomo nel suo complesso.

Per visualizzare l'idea di Bohr con un'immagine intuitiva potete pensare a una scala dove ogni gradino rappresenta uno stato atomico. La chiameremo la *"scala dell'energia",* perché ogni stato/gradino è contraddistinto da un valore determinato di questa grandezza fisica. Ciascun gradino corrisponde cioè ad un *"livello energetico",* intendendo con questo termine il valore dell'energia in uno degli stati permessi.

Normalmente un atomo può rimanere per un tempo indefinito ai piedi della scala. Quando si trova in questa condizione si dice che l'atomo è nello *"stato fondamentale",* caratterizzato dalla minima energia possibile. Assorbendo energia l'atomo può saltare su un livello/gradino superiore e in questo caso si dice che l'atomo passa in uno stato *"eccitato"*. Da uno stato eccitato l'atomo può tornare nello stato fondamentale (o comunque su un livello/gradino inferiore) cedendo energia e si parla allora di *"decadimento"* dell'atomo. Poiché il decadimento può avvenire spontaneamente, cioè senza assorbimento di energia, si verifica più facilmente dell'eccitazione, tranne quando l'atomo si trova già nello stato fondamentale, al di sotto del quale non può ulteriormente decadere. Se assorbe una quantità di energia superiore ad una soglia critica, l'atomo salta di botto tutti i gradini portandosi in cima alla scala, e di lì scivola via come su una superficie liscia. Quest'ultimo processo viene chiamato *"ionizzazione"* dell'atomo, perché lo trasforma in uno ione portandogli via un elettrone.

Abbiamo così l'immagine di un atomo che saltella su e giù per i gradini della scala dell'energia, assorbendo o cedendo una quantità di energia pari al dislivello coperto nel salto. La differenza di energia tra due livelli successivi, cioè l'altezza dei gradini, diminuisce man mano che si sale lungo la scala, come potete vedere nella figura (12) in fondo all'articolo. Noterete inoltre che i gradini divengono sempre più numerosi avvicinandosi alla cima della scala. In effetti, anche osservando col massimo ingrandimento non potreste contarli... ce ne sono un'infinità!

L'immagine della scala atomica dell'energia ci da un'idea intuitiva dei livelli energetici nell'atomo, che ci permette d'interpretare in modo elementare molti fenomeni non spiegabili con la sola fisica classica. È un'immagine al tempo stesso semplice, potente e pervasiva. In questo articolo seguiremo il filo logico-cronologico dei principali avvenimenti e scoperte che hanno portato alla luce la scala atomica dell'energia, dalle ipotesi ottocentesche di un legame tra spettri luminosi e struttura atomica, ai modelli atomici "classici" nel primo Novecento, fino alla teoria quantistica dell'atomo (1913). In particolare vedremo come dallo studio fenomenologico degli spettri atomici è stata dedotta l'esistenza di livelli energetici discreti e come questi sono stati visti *"entrare in azione"* quando gli atomi percorrono la scala dell'energia, negli esperimenti di collisione anelastica con elettroni (1914).

## *I - Gli spettri e la spettroscopia*

In fisica s'intende con il termine "spettro" la distribuzione dell'intensità della luce, e in generale di una radiazione, in funzione della lunghezza d'onda, della frequenza o dell'energia (o di qualche altra grandezza collegata). Si osservano spettri d'emissione e d'assorbimento. Gli spettri d'emissione si ottengono analizzando in frequenza le radiazioni emesse da sostanze eccitate (per riscaldamento passaggio di corrente elettrica, ecc). Per ottenere uno spettro d'assorbimento s'interpone tra una sorgente - che emette uno spettro continuo - e un analizzatore (un prisma o, meglio, un reticolo di diffrazione) una sostanza che assorbe selettivamente certe frequenze (o bande di frequenza). Inoltre si trova che le radiazioni assorbite dalla sostanza interposta sono esattamente le stesse che quella stessa sostanza sarebbe capace di emettere quando viene eccitata. Questo fenomeno viene chiamato inversione delle righe spettrali.

Gli spettri possono avere una struttura continua, a righe o a bande. Gli spettri continui sono emessi dai solidi e dai liquidi portati ad alta temperatura (e in certe condizioni anche dai gas), mentre gli spettri a righe sono tipici di atomi e ioni in fase gassosa a bassa pressione; gli spettri a bande, infine sono caratteristici di molecole e sostanze composte. Uno spettro con una sola riga corrisponde ad una radiazione monocromatica, cioè di un'unica frequenza. La registrazione di uno spettro su una lastra fotografica prende il nome di spettrogramma, e l'analisi degli spettrogrammi forma l'oggetto della spettroscopia.

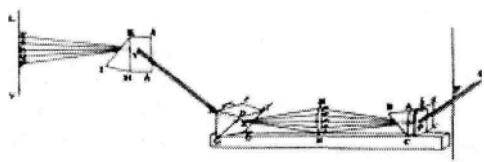

figura (1): scomposizione della luce coi prismi di Newton

Le prime indagini sugli spettri ottici si fanno risalire alle celebri esperienze di Newton sulla scomposizione della luce solare coi prismi (1666), di cui vedete un'illustrazione nella figura (1), tratta dall'edizione del 1721 *dell'Ottica*. Dopo Newton e durante tutto il Settecento la spettroscopia non fece grandi progressi, e soltanto all'inizio dell'Ottocento vennero avviate ricerche sistematiche. Dalle iniziali osservazioni sulla luce solare si passò all'analisi degli spettri di corpi celesti, fiamme, sostanze chimiche termosensibili. Uno dei risultati più importanti ottenuti in quel periodo - con l'impiego di nuove sorgenti luminose e strumenti di grande precisione - fu la scoperta che lo spettro ottico si estende al di là della regione visibile (1800/1802).

Nel corso dell'Ottocento si comprese che le tecniche spettroscopiche, consentendo agli scienziati di raccogliere un'enorme messe d'informazioni sulla struttura della materia, costituivano un formidabile metodo d'indagine in fisica, chimica e astronomia. Nel 1859 Bunsen aveva dimostrato che l'esistenza di una data riga in uno spettro indica sempre la presenza di un dato elemento. Questa scoperta, unita all'osservazione che la radiazione monocromatica è ottenuta solo dai gas atomici rarefatti - sistemi nei quali si può presumere che le interazioni tra atomi siano trascurabili e pertanto, come oggi sappiamo, i livelli energetici non sono perturbati e le righe spettrali sono ben risolte - rafforzò l'idea, suggerita da Stokes alcuni anni prima (1852), che le righe degli spettri d'emissione fossero dovute a vibrazioni atomiche di natura meccanica, della stessa frequenza della luce emessa, mentre l'assorbimento generava vibrazioni atomiche della stessa frequenza della luce assorbita, in accordo con il fenomeno meccanico della risonanza[1].

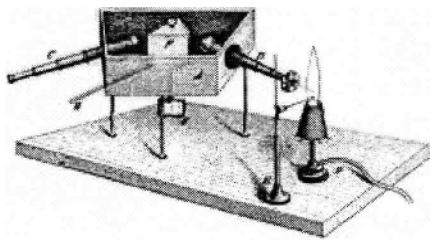

figura (2): spettroscopio di Kirchhoff e Bunsen

Come corollario di quest'idea, un gas di atomi che possono vibrare in più modi dà origine a uno spettro a righe - con tante righe quanti sono i modi vibratori possibili - ciascuna d'intensità proporzionale alla popolazione atomica che vibra alla frequenza corrispondente. Viceversa, una radiazione di una data frequenza eccita un atomo a vibrare nel modo corrispondente, con la comparsa di una riga d'assorbimento per ogni frequenza in grado di eccitare un modo di vibrazione presente nella popolazione atomica. La reciprocità fra spettri d'emissione e d'assorbimento, che era stata ipotizzata da Ångström nel 1853, fu confermata nel 1860 dall'osservazione da parte di Kirchhoff e Bunsen del fenomeno dell'inversione delle righe spettrali.

---

[1] secondo le diverse interpretazioni a vibrare era l'intera struttura atomica oppure le parti mobili all'interno dell'atomo.

## II—Lo spettro dell'idrogeno atomico

Se gli spettri portano informazioni sulla struttura atomica, dovremo ricavare molte informazioni dagli spettri più complessi. Ma anche il più semplice, quello dell'idrogeno allo stato atomico, quasi una "stele di Rosetta" della fisica atomica, permette di ricavare informazioni fondamentali.

Lo spettro d'emissione del gas d'idrogeno, contenuto in un tubo di Plücker[2] attraversato da una corrente elettrica, è uno spettro a molte righe, caratterizzate da bassa luminosità, elevata densità spettrale e struttura complessa. Aumentando la corrente nel tubo compare per contrasto un secondo spettro con caratteristiche complementari (intensità elevata, bassa densità spettrale e struttura semplice). L'intensità di questo secondo spettro è proporzionale alla corrente, mentre l'intensità dello spettro di fondo a molte righe aumenta solo lentamente quando viene aumentata la corrente nel tubo. Il fondo a molte righe è associato alla radiazione emessa da molecole d'idrogeno ($H_2$), mentre lo spettro in contrasto è dovuto ad atomi d'idrogeno (H) prodotti per dissociazione di molecole indotta dalla corrente.

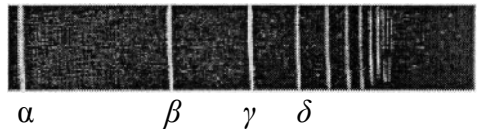

figura (3): spettro dell'idrogeno atomico

Le prime tre righe della regione visibile dello spettro dell'idrogeno atomico, mostrato nella figura (3), furono individuate nel 1862 da Ångström, che ne trovò poi una quarta nel 1871 e determinò le seguenti lunghezze d'onda (in Å $=10^{-8}$cm):[3]

$$\lambda_\alpha=6562.10,$$
$$\lambda_\beta=4860.74,$$
$$\lambda_\gamma=4340.10,$$
$$\lambda_\delta=4101.20.$$

Stoney notò nel 1871 che la prima, seconda e quarta riga di Ångström erano inversamente proporzionali, rispettivamente, ai numeri *20, 27* e *32*. Ne concluse che erano la *20esima, 27esima* e *32esima* armonica di una vibrazione fondamentale, come possiamo vedere se ne calcoliamo i rapporti (arrotondati alla terza cifra decimale):

$$\frac{\lambda_\alpha}{\lambda_\beta} = 1.350021 \simeq 1.35 = \frac{27}{20},$$

$$\frac{\lambda_\alpha}{\lambda_\gamma} = 1.511970 \simeq 1.512 = \frac{189}{125} = \frac{9 \cdot 21}{5 \cdot 25},$$

$$\frac{\lambda_\alpha}{\lambda_\delta} = 1.600044 \simeq 1.6 = \frac{8}{5} = \frac{32}{20}.$$

---

[2] un tubo di Plücker (o di Geissler) è un tubo di vetro a forma della lettera H, costituito da due bulbi collegati da un sottile capillare. Due elettrodi collocati alle estremità del tubo provocano una scarica elettrica nel gas contenuto all'interno. La scarica nel capillare può avvenire a bassa pressione e a temperatura ambiente. In queste condizioni la riga emessa è molto netta e luminosa.

[3] le misure moderne forniscono questi valori: $\lambda_\alpha$=6562.8, $\lambda_\beta$=4861.3, $\lambda_\gamma$=4340.5, $\lambda_\delta$=4101.7.

Come si vede, la prima e l'ultima frazione sono abbastanza semplici, mentre quella intermedia è più complicata. Proviamo ora ad esprimere le misure di Ångström come frazioni di una lunghezza fondamentale $\lambda$. Avremo:

$$\lambda_\alpha = r_\alpha \lambda, \quad \lambda_\beta = r_\beta \lambda, \quad \lambda_\gamma = r_\gamma \lambda, \quad \lambda_\delta = r_\delta \lambda,$$

ovvero $\quad r_\beta = r_\alpha \dfrac{20}{27}, \quad r_\gamma = r_\alpha \dfrac{5 \cdot 25}{9 \cdot 21}, \quad r_\delta = r_\alpha \dfrac{5}{8}.$

Poiché contano solo i rapporti tra le lunghezze, $r_\alpha$ si può scegliere a piacere. Scegliendo per comodità $r_\alpha = \dfrac{9}{5}$ otteniamo: $r_\beta = \dfrac{9}{5} \cdot \dfrac{20}{27} = \dfrac{4}{3}, \quad r_\gamma = \dfrac{9}{5} \cdot \dfrac{5 \cdot 25}{9 \cdot 21} = \dfrac{25}{21} =, \quad r_\delta = \dfrac{9}{5} \cdot \dfrac{5}{8} = \dfrac{9}{8}.$

Nella sequenza così ottenuta: $\dfrac{9}{5}, \dfrac{4}{3}, \dfrac{25}{21}, \dfrac{9}{8}$, ai numeratori ci sono quadrati perfetti. Se facciamo in modo che i quadrati ai numeratori formino una progressione geometrica la sequenza diventa: $\dfrac{9}{5}, \dfrac{16}{12}, \dfrac{25}{21}, \dfrac{36}{32}$, dove potete notare che i denominatori si ottengono sottraendo 4 dai numeratori. Perciò possiamo rappresentare le misure di Ångström mediante la serie $\dfrac{3^2}{3^2 - 4}, \dfrac{4^2}{4^2 - 4}, \dfrac{5^2}{5^2 - 4}, \dfrac{6^2}{6^2 - 4}.$

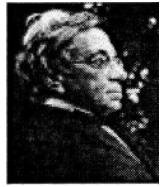

**J.J. Balmer
(1825-1898)**

Questo semplice ragionamento numerico ci guida alla formula che fu trovata da Balmer nel 1885:

$$(1) \quad \lambda_n = \dfrac{n^2}{n^2 - 2^2} \lambda_\infty, \quad n = 3, 4, 5, ...,$$

dove $\lambda_\infty$ è il limite della serie quando *n* tende all'infinito. Questa riga limite si può determinare con ottima approssimazione da $(r_\alpha + r_\beta + r_\gamma + r_\delta)\lambda_\infty \simeq (\lambda_\alpha + \lambda_\beta + \lambda_\gamma + \lambda_\delta)$, ottenendo $\lambda_\infty \simeq 3645.6 \text{Å}$ [4].

All'epoca della sua scoperta la (1) era considerata una strana formula, perché ci si attendeva come per le onde sonore una dipendenza armonica, cioè una frazione semplice del tipo *m/n* di una lunghezza fondamentale, senza termini al quadrato.

La formula di Balmer (1) descrive con grande precisione le righe dell'omonima serie, rappresentate schematicamente nella figura (4). Come si vede, le righe si addensano dalla parte delle lunghezze d'onda decrescenti (frequenze crescenti) fino al limite della serie $\lambda_\infty$, che si trova nel vicino ultravioletto (NUV). Quasi tutte le righe Balmer si trovano nel NUV; fanno eccezione le prime cinque (nella figura ne sono indicate 4: $H_\alpha$, $H_\beta$, $H_\gamma$, $H_\delta$), rivelate nel visibile.

---

[4] e numero d'onda limite $\tilde\nu_\infty \simeq 27419.5 \text{ cm}^{-1}$. Le lunghezze delle righe di Ångström calcolate dalla (1) con $\lambda_\infty \simeq 3645.6 \text{Å}$ sono: $\lambda_\alpha = 6562.08 \text{Å}$, $\lambda_\beta = 4860.8 \text{Å}$, $\lambda_\gamma = 4340 \text{Å}$, $\lambda_\delta = 4101.3 \text{Å}$, che si discostano di appena $0.1 \text{ Å}$ dai valori misurati. Sono state individuate, e accuratamente misurate, una trentina di righe della serie Balmer. Utilizzando le misure moderne, la coincidenza tra valori misurati e calcolati risulta pressoché perfetta.

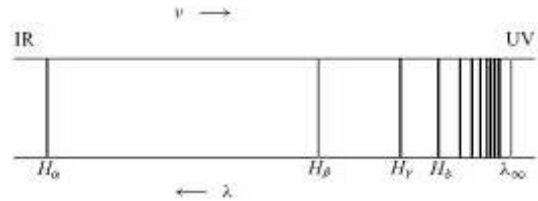

figura (4): serie di Balmer

L'individuazione nelle regioni UV (ultravioletto) e IR (infrarosso), a partire dal 1904, di nuove serie che non si accordavano con la (1), dimostrò però che la formula di Balmer non era in grado di descrivere l'intero spettro dell'idrogeno.

In effetti, una formula molto più generale, espressa in funzione del numero d'onda $\tilde{\nu} = \dfrac{1}{\lambda}$, era stata trovata da Rydberg nel 1890[5]. Nel caso dell'idrogeno, la formula di Rydberg può essere vista come una diretta generalizzazione della formula di Balmer (1). Se esprimiamo quest'ultima in termini di numeri d'onda avremo:

$$\frac{1}{\lambda_n} \equiv \tilde{\nu}_n = \frac{(n^2 - 2^2)}{n^2} \cdot \frac{1}{\lambda_\infty} = \left(1 - \frac{4}{n^2}\right)\frac{1}{\lambda_\infty} = \frac{4}{\lambda_\infty}\left(\frac{1}{2^2} - \frac{1}{n^2}\right)$$, da cui troviamo

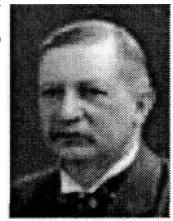

**J. Rydberg (1854-1919)**

$$(1') \quad \tilde{\nu}_n = R\left(\frac{1}{2^2} - \frac{1}{n^2}\right),\ [6]$$

dove $R = \dfrac{4}{\lambda_\infty}$. La $(1')$ esprime la formula di Balmer come caso particolare della legge

$$(2) \quad \tilde{\nu} = R\left(\frac{1}{n'^2} - \frac{1}{n^2}\right).$$

La (2) è una legge importantissima che fu introdotta da Ritz per la prima volta nel 1903, e per questo motivo è nota come legge di Rydberg-Ritz per l'idrogeno[7]; in essa $n = 1, 2, 3, 4, \ldots$ indica il numero d'ordine della serie spettrale, mentre $n = n'+1$, $n'+2$, $n'+3$, ... denota il numero della riga, appartenente alla serie d'ordine *n'*. La grandezza *R* prende il nome di costante di Rydberg e vale, per l'idrogeno[8], $R_H = 109678 \text{ cm}^{-1}$.

---

[5] una formula alternativa del tipo $\tilde{\nu} = a + \dfrac{b}{n^2} + \dfrac{c}{n^4} + \ldots$ fu proposta anch'essa intorno al 1890 da Kayser e Runge con qualche successo, ma oggi rimane una mera curiosità storica; la formula di Rydberg si dimostrò quella giusta in virtù dell'interpretazione teorica datane da Bohr (1913).

[6] e frequenza corrispondente $\nu_n = \left(\dfrac{1}{4} - \dfrac{1}{n^2}\right) \cdot 3.29 \cdot 10^{15}\,\text{Hz}$.

[7] la formula originaria di Rydberg era ancor più generale della (2), non essendo limitata al solo caso dell'idrogeno.

La legge di Rydberg-Ritz (2) descrive molto bene lo spettro dell'idrogeno in tutte le regioni esplorate:

- per *n'=1* e *n=2, 3, 4, ...* riproduce la serie di Lyman (1904-06) nella regione UV;

- per *n'=2* e *n=3, 4, ...* si riduce alla (1') già nota per la serie di Balmer;

- infine, per *n'=3* (*n=4, 5, ...*), *n'=4* (*n=5, 6, ...*), *n'=5* (*n=6, 7, ...*) ha fornito la prima previsione teorica (Ritz, 1908) delle serie IR poi scoperte rispettivamente da Paschen (1908), Brackett (1922) e Pfund (1924).

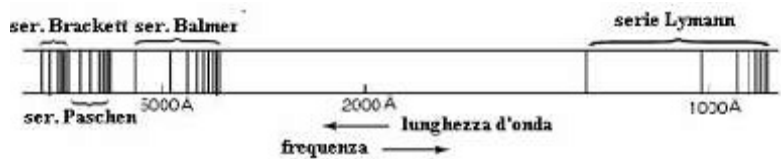

figura (5): serie spettrali dell'idrogeno

Quando affermiamo che la (2) fu in grado di fornire previsioni teoriche di nuove serie spettrali ci riferiamo al fatto che si tratta di una legge che fu dedotta da un modello teorico generale dell'atomo (Ritz, 1908), e non semplicemente ricavata dai dati sperimentali, come nel caso della (1) di Balmer. Anche se il modello di Ritz si dimostrò errato, la successiva corretta deduzione della (2) da parte di Bohr (1913) le conferì definitivamente la dignità di una vera legge fisica, e come tale naturalmente va considerata retrospettivamente.

La legge di Rydberg-Ritz (2) si applica altrettanto bene agli spettri degli ioni idrogenoidi, che come l'idrogeno hanno un solo elettrone in orbita attorno al nucleo (di carica positiva *Z*). Si tratta di ioni positivi ottenuti ionizzando *Z-1* volte atomi di numero atomico Z, come per esempio $He^+$ (Z=2), $Li^{++}$ (Z=3), $Be^{+3}$ (Z=4) che hanno una carica elettrica pari a *Z-1* volte quella del protone. Per questi ioni basta semplicemente moltiplicare per $Z^2$ il corrispondente numero d'onda dell'idrogeno:

$$(3) \tilde{\nu} = Z^2 R \left( \frac{1}{n'^2} - \frac{1}{n^2} \right).$$

Osserviamo ora un fatto curioso. Se consideriamo lo ione $He^+$ (Z=2), per *n'=4* e *n=5, 6, ...* otteniamo la serie di Pickering,[9] che ha la notevole peculiarità di coincidere[10] con la serie Balmer per *n* pari, ma ha un numero doppio di righe. Curiosamente, le righe della serie Pickering furono individuate per la prima volta nello spettro di una stella (1896) ed interpretate come righe Balmer[11].

---

[8] *R* è una costante per le diverse serie spettrali di una stessa specie atomica, ma cambia leggermente da una specie all'altra.

[9] $\tilde{\nu} = R \left( \frac{1}{4} - \frac{1}{(n/2)^2} \right)$ e in generale per lo ione di carica +(Z-1): $\tilde{\nu} = R \left( \frac{1}{4} - \frac{1}{(n/Z)^2} \right)$ per *n'=2Z* e *n=n'+1, n'+2,...* .

[10] non si tratta di una coincidenza perfetta perché *R* cambia leggermente da un elemento all'altro.

[11] Rydberg trovò una piccola differenza tra il valore misurato della riga 4686 Å e il valore calcolato 4688 Å ma l'attribuì ad un errore nelle osservazioni.

Perciò le righe dell'elio ionizzato furono attribuite inizialmente all'idrogeno atomico e solo molto più tardi (Bohr, 1913) vennero identificate correttamente.

### *III — Il principio di combinazione*

Il risultato fondamentale delle ricerche spettroscopiche dell'Ottocento è stata la scoperta di regolarità spettrali racchiuse in leggi generali, come quelle di Balmer o di Rydberg. Queste regolarità non erano limitate allo spettro dell'idrogeno. Si scoprì, per esempio, che le righe spettrali degli elementi alcalini potevano essere raggruppate in sequenze (o serie spettrali), descritte da leggi "differenziali" del tipo (2) di Rydberg-Ritz.[12]

Ancor più in generale, negli spettri di atomi, ioni e molecole, le righe di una data sequenza, pur presentando caratteristiche differenti in intensità, larghezza, struttura (fina o iperfina), comportamento in un campo elettrico o magnetico (effetti Stark, Zeeman e Paschen-Back), ecc., hanno tutte la stessa dipendenza dal numero d'onda[13]. Questa dipendenza si può esprimere come la differenza di due termini, detti "termini spettroscopici", il primo dei quali è una costante per la serie spettrale data (limite della serie), mentre il secondo cambia da riga a riga determinando una successione in corrispondenza ai numeri interi a partire da un intero fissato. La successione di termini così formata non va confusa con la corrispondente sequenza di righe spettrali.

In generale i termini spettroscopici non hanno la semplice forma $\frac{R}{n^2}$ che contraddistingue l'idrogeno. I termini spettroscopici di Rydberg, che descrivono bene molte serie, sono definiti dall'espressione $\tau_n = \frac{Z^2 R}{(n-\alpha)^2}$, dove $\alpha$ (chiamata "correzione di Rydberg") è una costante caratteristica della successione. Come noterete immediatamente, per $\alpha = 0$ ritroviamo i

---

[12] Si tratta delle serie seguenti:

- "principale" $\qquad \tilde{\nu} = R\left(\dfrac{1}{(1+s)^2} - \dfrac{1}{(n+p)^2}\right), n = 2, 3, ...,$

- "diffusa" $\qquad \tilde{\nu} = R\left(\dfrac{1}{(2+p)^2} - \dfrac{1}{(n+d)^2}\right), n = 3, 4, ...,$

- "fine" $\qquad \tilde{\nu} = R\left(\dfrac{1}{(2+p)^2} - \dfrac{1}{(n+s)^2}\right), n = 2, 3, ...,$

. "fondamentale" $\qquad \tilde{\nu} = R\left(\dfrac{1}{(3+d)^2} - \dfrac{1}{(n+f)^2}\right), n = 4, 5, ...,$

ecc., dove le costanti *s, p, d, f,* ecc. sono numeri frazionari che definiscono ciascuna serie.

[13] quando parliamo di numero d'onda ci riferiamo al numero d'onda spettroscopico $\tilde{\nu}$ (indicato in letteratura anche con la lettera $\kappa$), definito da $\tilde{\nu} = \kappa = \dfrac{1}{\lambda_0} = \dfrac{\nu}{c}$, dove $\lambda_0$ è la lunghezza d'onda misurata nel vuoto. In fisica atomica si usano anche il numero d'onda nel vuoto $\kappa_0 = 2\pi\kappa = \dfrac{2\pi}{\lambda_0} = \dfrac{\omega}{c}$ e il numero d'onda $n\kappa_0 = \dfrac{2\pi}{\lambda} = n\dfrac{\omega}{c} = \dfrac{\omega}{v}$, dove $n$ è l'indice di rifrazione.

termini di Balmer $\frac{Z^2 R}{n^2}$ come caso particolare.

In molti casi gli spettri sono ancor meglio rappresentati dai termini spettroscopici di Ritz $\tau_n = \frac{Z^2 R}{(n-\delta_n)^2}$, dove la quantità $\delta_n$ prende il nome di "difetto quantico"[14]; $\alpha$ anche qui è la correzione di Rydberg, mentre $\beta$ ("correzione di Ritz") è un'altra costante caratteristica della successione (di regola le correzioni successive sono trascurabili). Si trova sperimentalmente che $\alpha$ è positiva e $\beta$ negativa, e che $\alpha$ è maggiore di $\beta$ in valore assoluto; inoltre il difetto quantico $\delta_n$ diviene sempre più prossimo ad $\alpha$ all'aumentare di $n$.

Ad un dato atomo corrispondono dunque, per diversi valori di $\alpha$ (ed eventualmente di $\beta$) diverse successioni di termini:

- $\tau_1, \quad \tau_2, \quad \tau_3, \quad ..., \quad \tau_n \qquad$ in corrispondenza di $\alpha(\beta)$,

- $\tau'_1, \quad \tau'_2, \quad \tau'_3, \quad ..., \quad \tau'_n \qquad$ in corrispondenza di $\alpha'(\beta')$,

- $\tau''_1, \quad \tau''_2, \quad \tau''_3, \quad ..., \quad \tau''_n \qquad$ in corrispondenza di $\alpha''(\beta'')$,

- ecc.

Di regola, il numero d'onda di una riga si può esprimere, come abbiamo detto, come differenza di due termini spettroscopici appartenenti a due diverse successioni:

$$(4) \quad \tilde{\nu}_{n'n} = \tau'_{n'} - \tau_n,$$

dove $\tilde{\nu}_{n'n}$ rappresenta il numero d'onda dell'*nesima* riga della serie di ordine $n'$. Per esempio il numero d'onda delle righe descritte dai termini di Rydberg è dato da $\tilde{\nu} = Z^2 R \left( \frac{1}{(n'-\alpha')^2} - \frac{1}{(n-\alpha)^2} \right)$, che si riduce alla (3) quando $\alpha=\alpha'=0$.

La (4) afferma che un numero relativamente limitato di termini spettroscopici può produrre un grande numero di righe: precisamente, $n$ termini posson dar luogo a $n(n-1)/2$ righe.

Rydberg aveva trovato sperimentalmente che la riga limite di una serie (per esempio $\tilde{\nu}_{n'\infty} = \tau'_{n'}$ per la serie di ordine $n'$) può coincidere numericamente con un termine appartenente ad un'altra serie (per esempio la serie di ordine $n''$). Quando questo avviene vi saranno righe appartenenti a quest'ultima serie che hanno il numero d'onda dato da

---

[14] $\delta_n = \alpha + \frac{\beta}{(n-\delta_n)^2} + \frac{\gamma}{(n-\delta_n)^4} + ...$ si ottiene da $\tau_n = \frac{Z^2 R}{(n-\delta_n)^2}$ definendo $t_n \equiv \frac{1}{(n-\delta_n)^2}$. Allora si può sviluppare $\delta_n = n - \frac{1}{\sqrt{t_n}}$ in serie di $t_n$ ottenendo $\delta_n = \alpha + \beta t_n + \gamma t_n^2 + ...$, dove i coefficienti $\alpha, \beta, \gamma, ...$ si possono determinare numericamente riportando su un grafico le misure di $\delta_n$ e $t_n$. Uno sviluppo in serie di questo tipo è noto come espansione di Ritz (1903).

$$(5)\, \tilde{v}_{n''n'} = \tau''_{n''} - \tau'_{n'}.$$

Sommando la (4) e la (5) si trova l'importantissima relazione:

$$(6)\, \tilde{v}_{n''n} = \tilde{v}_{n''n'} + \tilde{v}_{n'n}$$

dove ora $\tilde{v}_{n''n}$ è il numero d'onda dell'*nesima* riga appartenente alla serie di ordine *n''*, e ovviamente vale anche

$$(6')\, \tilde{v}_{n'n} = \tilde{v}_{n''n} - \tilde{v}_{n''n'}.$$

Le (6) e (6') esprimono il "principio di combinazione di Ritz": *i numeri d'onda di una serie spettrale si possono esprimere come somme (o differenze) di un più limitato numero di altri numeri d'onda*. Il principio di combinazione consente di esprimere nuove righe mediante quelle note, combinandole per somma o per differenza. Si noti però che non si tratta di una regola di validità assoluta, perché non sempre (anche se molto spesso) la somma (o la differenza) di due numeri d'onda di un atomo è uguale ad un altro numero d'onda dello stesso atomo. Il principio di combinazione è il principio guida per la comprensione teorica degli spettri a righe; esso costituì la base fenomenologica della relazione energia-frequenza di Bohr (15) e, in seguito, della meccanica matriciale di Heisenberg.

### *IV— Stabilità e dimensioni degli atomi*

Agli inizi del Novecento la spiegazione delle regolarità racchiuse nelle leggi di Balmer e di Rydberg-Ritz costituiva uno dei principali problemi irrisolti della fisica dell'epoca. Per la verità, non soltanto le regolarità espresse dalle leggi spettrali ma l'esistenza stessa di righe nette e definite come quelle che venivano osservate negli spettri dei gas rimaneva enigmatica. I modelli *"classici"* dell'atomo, come quello di Thomson (1904) o *l'atomo nucleare* di Rutherford (1911) - tralasciando i meno noti - erano incapaci di fornire un'interpretazione teorica dei risultati sperimentali della spettroscopia.

Questi primi modelli atomici quantitativi degli inizi del Novecento nacquero invero alquanto tardivamente. In precedenza le leggi spettroscopiche, scoperte empiricamente, non erano state sottoposte a test teorici, in mancanza di un'elettrodinamica in grado di collegarle in modo coerente con i moti vibratori degli atomi (o di parti di essi) che si presumeva ne fossero all'origine. Una teoria di questo genere fu disponibile soltanto dopo il 1895, anno in cui Lorentz formulò una teoria degli elettroni fondata sull'elettromagnetismo maxwelliano. Forse le ragioni di questo ritardo risiedono nello status delle teorie maxwelliane del campo elettromagnetico (1865) e della luce (1873), che solo nel 1888, con l'inequivocabile conferma sperimentale ottenuta da Hertz, acquisirono quella solida base empirica che loro mancava. Fu infine la conferma della corretta previsione del rapporto *e/m* per l'elettrone, ottenuta da Zeeman (1896) studiando lo sdoppiamento delle righe spettrali nei campi magnetici (effetto Zeeman), a convincere i fisici che l'elettrodinamica classica (o elettrodinamica di Larmor-Lorentz in onore dei due scienziati che diedero i maggiori contributi a questa disciplina) poteva essere applicata alla teoria atomica.

Tuttavia l'elettrodinamica di Larmor-Lorentz, se riusciva a interpretare a grandi linee i processi di radiazione in termini di movimenti di elettroni, mal si conciliava con l'evidenza dei caratteristici spettri a righe, e tanto meno riusciva a descriverne le proprietà in modo particolareggiato. Infatti le leggi della meccanica e dell'elettrodinamica classica, quando vengono applicate ad oscillatori carichi come quelli che si supponevano presenti nell'atomo, portano alle seguenti previsioni:

1. la diminuzione dell'ampiezza di oscillazione delle cariche, dovuta all'irraggiamento di onde elettromagnetiche da parte di queste ultime;
2. la proporzionalità dell'intensità della luce irradiata alla quarta potenza della frequenza meccanica dell'oscillatore.

Se si accettano tali previsioni l'atomo "classico", concepito come un sistema isolato contenente un oscillatore carico, non può essere stabile, ma deve decadere spontaneamente per irraggiamento di onde elettromagnetiche. Il calcolo preciso della durata media del processo d'irraggiamento, eseguito con le leggi elettrodinamiche di Larmor-Lorentz nell'approssimazione di dipolo elettrico atomico, prevede un tempo di decadimento dell'ordine del centomilionesimo di secondo ($10^{-8}$s). Il lettore interessato potrà trovare questo calcolo nell'*Appendice I*. In questo processo verrebbero irradiate onde di tutte le frequenze, con un picco centrale in corrispondenza della frequenza naturale dell'oscillatore. Di conseguenza, le righe spettrali non potrebbero essere sottili e ben marcate come si osserva, ma sfumate (spettro continuo), e con un caratteristico allargamento (curva lorentziana, figura (6)).

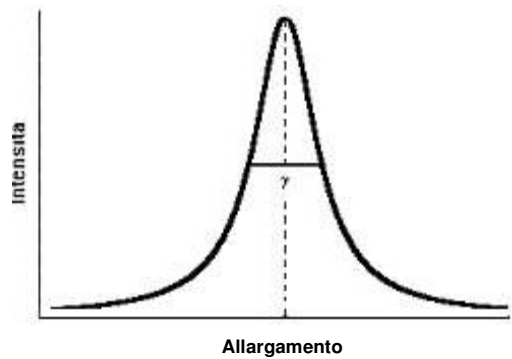

figura (6): curva lorentziana

Per comprendere intuitivamente il problema potete considerare l'atomo "classico" come una minuscola antenna dipolare. Questo vale se si immagina, come nel modello di Thomson del 1904, che gli elettroni siano disposti in anelli concentrici all'interno di una sfera uniforme di elettricità positiva. Ciascun elettrone descrive un'orbita lungo l'anello e contemporaneamente può oscillare attorno a una posizione di equilibrio dinamico, producendo il segnale oscillante che fa irradiare l'antenna. Nell'*atomo nucleare*, in cui gli elettroni orbitano attorno ad un piccolo e compatto nucleo positivo, il moto dell'elettrone si può descrivere come una combinazione di due oscillazioni perpendicolari. Di nuovo, l'atomo si comporta come un'antenna radiante. In entrambi i casi la potenza irradiata è proporzionale al quadrato dell'accelerazione dell'elettrone (legge di Larmor); di conseguenza l'atomo perde energia e (se non riceve nuova energia dall'esterno) il movimento degli elettroni cessa rapidamente.

In alcuni casi, però, la perdita d'energia per radiazione può essere molto bassa. Se vi sono *n* cariche tutte uguali, per esempio *n* elettroni, l'accelerazione dell'elettrone nella legge di Larmor dev'essere sostituita dalla somma vettoriale delle accelerazioni di tutti gli elettroni. Poiché si tratta di una somma vettoriale, se vi sono in media tanti vettori orientati in una direzione quanti nella direzione opposta, è possibile che la somma si annulli in ogni istante. Affinché ciò si verifichi è sufficiente che il momento di dipolo elettrico del sistema sia nullo. Thomson, nel suo modello atomico del 1904, determinò appunto le condizioni per una distribuzione di cariche con momento dipolare elettrico nullo, ben sapendo che in questa situazione la perdita d'energia del sistema per radiazione sarebbe stata bassissima (nessuna radiazione di dipolo elettrico, pochissima radiazione di quadripolo elettrico e dipolo magnetico). I calcoli inoltre dimostravano che le perdite si sarebbero ulteriormente ridotte all'aumentare del numero di cariche del sistema. Era dunque sufficiente aggiungere semplicemente un numero sempre maggiore di elettroni nelle configurazioni a bassa emissione per stabilizzare un atomo. Questa non era una difficoltà perché all'epoca si riteneva che un atomo contenesse migliaia di elettroni. In questo modo un atomo poteva raggiungere una vita media estremamente lunga. Trovava spiegazione anche la radioattività, attribuita all'instabilità che si manifestava quando la velocità degli elettroni orbitanti scendeva al di sotto di un valore critico.

Un problema di meno agevole soluzione era quello delle dimensioni dell'atomo. Nei modelli atomici classici si doveva infatti introdurre artificialmente un parametro delle dimensioni di una lunghezza, come il raggio della sfera positiva di Thomson o il "raggio dello ione" di Jeans, che forniva la scala lineare delle dimensioni atomiche. In mancanza di tale parametro non vi sarebbe limite, secondo la fisica classica, alle dimensioni degli atomi - cosa palesemente assurda. Se le cariche elementari (elettroni o "ioni") vengono considerate puntiformi, come richiede l'elettrodinamica, non si può in alcun modo ottenere una lunghezza partendo dalle sole grandezze permanentemente associate agli elettroni (numero, massa e carica). L'inserimento *ad hoc* di un parametro dimensionale nei modelli atomici poteva risolvere il problema ma ne poneva altri; per esempio, perché la sfera positiva di Thomson, come ogni altro corpo carico esteso, non avrebbe dovuto espandersi indefinitamente per la reciproca repulsione delle parti?

Ma il problema più grave e difficile rimaneva quello dell'interpretazione teorica degli spettri: nessuno riusciva a ricavare, partendo dai parametri di un modello teorico dell'atomo, la formula di Rydberg.

## *V — L'atomo magnetico*

La ricerca di un meccanismo di spiegazione delle regolarità spettrali sembrava inevitabilmente incontrare difficoltà insormontabili. In primo luogo, le equazioni del modello dell'elettrone legato elasticamente (o quasi-elasticamente) ammettono delle soluzioni (chiamate autovalori) in cui figura il quadrato della frequenza, mentre le leggi spettroscopiche presentano una semplice dipendenza lineare dalla frequenza. In secondo luogo, queste soluzioni non prevedono un limite della serie, come invece si osserva.

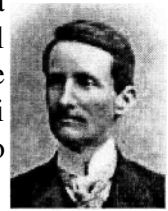

**W. Ritz
(1878-1909)**

Questa era la situazione, apparentemente senza via d'uscita, cui Ritz si trovava di fronte quando intraprese il suo tentativo di trovare una descrizione teorica delle regolarità spettrali nel quadro della fisica classica. Ritz era un fisico teorico di prim'ordine, capace di utilizzare al meglio gli strumenti più avanzati e potenti della sua disciplina[15]. Aveva cominciato considerando un modello elastico dell'atomo, immaginato come una membrana bidimensionale flessibile, ed ottenendo per le corrispondenti equazioni una serie di autovalori del tipo $\nu^2 \propto \left(n'^2 + n^2\right)$, dove $n'$, $n$ sono due numeri interi. Questo risultato aveva tre conseguenze in stridente contrasto con l'esperienza:

---

[15] un'interessante rassegna del lavoro e dei metodi di Ritz in spettroscopia teorica si trova in M. A. Elyashevich, N.G. Kembrovskaya, L.M. Tomil'chik. *Walter Ritz as a theoretical physicist and his research on atomic spectra theory,* in "Physics-Uspeckhi", 38 (4), 435-455, 1995. Sul contributo di Ritz alla teoria dell'elasticità segnaliamo V.V. Meleshko, *Selected topics in the history of the two-dimensional biharmonic problem,* in "Applied Mechanics Reviews", 56 (1), 33-85, 2003.

1. un aumento infinito della frequenza al crescere di $n'$, $n$, mentre le formule spettroscopiche tendono verso un limite finito della frequenza;

2. una dipendenza dal quadrato della frequenza anziché dalla prima potenza;

3. una formula simmetrica rispetto a $n'$, $n$, mentre la formula di Rydberg è antisimmetrica.

Ritz osò allora operare una modifica, per la verità alquanto artificiale, delle equazioni elastiche e riuscì a riprodurre la formula di Rydberg per grandi valori di $n'$, $n$. Desideroso di dare una base fisica alla sua ardita costruzione matematica, si rese conto che ciò richiedeva una modifica della definizione dell'energia potenziale elettromagnetica del sistema. Sfortunatamente ne risultava l'introduzione di una "forza" che non aveva un'interpretazione fisica ragionevole.

A questo punto Ritz abbandonò il modello puramente elastico della membrana bidimensionale e adottò un metodo più euristico. Osservò innanzitutto che per ottenere relazioni lineari nella frequenza $v$ e non nel quadrato $v^2$ si deve partire da equazioni differenziali del primo ordine, anziché del secondo come nei modelli elastici. Equazioni del primo ordine richiedono che le forze dipendano dalla velocità, come avviene per le forze magnetiche.

Sulla base di tali considerazioni Ritz elaborò un ingegnoso modello misto elastico-elettromagnetico, considerando un filamento elettricamente carico che ruotava intorno ad un asse generando un campo magnetico. Giunse in questo modo a ricavare una formula che aveva la stessa forma di quella di Rydberg. Ma Ritz non era ancora soddisfatto: doveva trovare una base fisica che rendesse meno artificiale il nuovo modello. Alla fine decise di rinunciare a qualsiasi elemento elastico e sostituì il filamento con un campo magnetico atomico. Questo modello magnetico dell'atomo (1908), oltre a fornire la prima interpretazione teorica della formula di Rydberg, dava anche una risposta al problema delle dimensioni atomiche, poiché introduceva in modo naturale un parametro dimensionale, cioè la lunghezza della calamita atomica generatrice del campo.

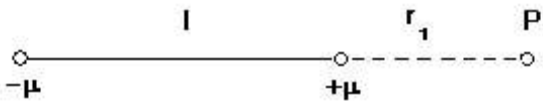 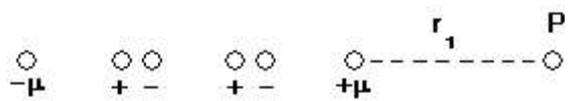

figura (7): calamita atomica    figura (8): catena di magneti elementari

Ritz congetturò quindi che l'atomo fosse la sede di un intenso campo magnetico generato da una microscopica ma potente calamita (o dipolo magnetico, figura (7)).

Il campo magnetico in un punto $P$ lungo l'asse della calamita è dato da $H = \mu \left[ \dfrac{1}{r_1^2} - \dfrac{1}{(l+r_1)^2} \right] = \mu \left( \dfrac{1}{r_1^2} - \dfrac{1}{r_2^2} \right)$, dove $r_1$ è la distanza di $P$ dal polo positivo e $l$ è la lunghezza della calamita. Esattamente lo stesso campo sarà creato dalla catena di magnetini mostrata nella figura (8). Se indichiamo con $d$ il numero dei magnetini elementari, ciascuno di lunghezza $a$ ($= \dfrac{l}{d}$), e se la distanza dal punto $P$ è $r_1 = n'a$, allora $r_2 = na$, dove $n = d+n'$, e vi sarà nel punto $P$ il campo

$(7) H = \dfrac{\mu}{a^2} \left( \dfrac{1}{n'^2} - \dfrac{1}{n^2} \right)$.

Ora, se nel punto *P* vi è un elettrone che può compiere piccoli movimenti in una ristretta area intorno a *P* nel piano perpendicolare all'asse del campo, questo elettrone si troverà immerso in un campo quasi omogeneo. L'intensità dell'ipotetico campo magnetico atomico, per le frequenze ottiche, era stimato dell'ordine di $10^8$ gauss, che era $10^4$ volte il campo esterno in cui venivano immersi gli atomi negli studi sull'effetto Zeeman.

Ritz ipotizzò che ogni riga spettrale fosse il risultato di una separazione per un effetto Zeeman di tipo particolare (cioè diverso dall'effetto Zeeman normale), causato - anche in assenza di un campo magnetico esterno - dall'azione del campo atomico interno. La separazione Zeeman in un campo magnetico - cioè la distanza (in frequenza) dalla posizione originale della riga spostata - è uguale alla frequenza di Larmor

$$(8) \nu_L = \frac{eH}{4\pi mc},$$

che rappresenta la frequenza del moto di precessione dell'elettrone in un campo magnetico uniforme (Larmor, 1900). L'interpretazione classica di questo spostamento è riportata nell'*Appendice II*.

Dalla (7) e dalla (8) si ottiene

$$(9) \tilde{\nu} = \frac{\mu e}{4\pi mc^2 a^2} \left( \frac{1}{n'^2} - \frac{1}{n^2} \right).$$

Se confrontiamo la (9) con la generalizzazione (2) della formula di Balmer troviamo l'espressione della costante di Rydberg $R = \frac{e}{4\pi mc^2} \cdot \frac{\mu}{a^2}$ in funzione dei parametri del modello, dove la lunghezza (*a*) e il numero (*d*) dei magneti elementari sono parametri variabili; la lunghezza assume carattere fondamentale, essendo una proprietà permanente dei magneti elementari.

È notevole come Ritz, introducendo le condizioni $r_1 = n'a$ e $r_2 = na$ per le distanze dell'elettrone dai due poli del magnete atomico, abbia fatto qualcosa che somiglia molto ad una quantizzazione della lunghezza: *distanze multiple di una lunghezza fondamentale danno luogo alle serie spettrali osservate*. Inoltre, variando la lunghezza di uno degli elementi estremi della catena, si poteva riprodurre la correzione di Rydberg; ed allo stesso modo si potevano ottenere i successivi termini correttivi (di Ritz, ecc.) variando le lunghezze del secondo magnetino (del terzo, ecc.). In definitiva, dipendendo dai parametri variabili del modello, ogni riga veniva a dipendere da un diverso stato dell'atomo.

Il modello di Ritz consentiva di dare un'interpretazione fisica molto semplice del principio di combinazione: ciascuno dei due termini nella (4) poteva essere interpretato come il contributo di uno dei poli della calamita atomica. Secondo questa interpretazione le differenti combinazioni dei termini spettrali erano semplicemente la conseguenza di possibili varianti nella posizione reciproca dei poli.

Nonostante questi successi, il modello magnetico dell'atomo non ebbe molta fortuna. Malgrado la sua sagacia Ritz non riuscì ad individuare un meccanismo fisico che spiegasse l'origine del campo atomico; fu comunque il primo a produrre previsioni teoriche di nuove serie spettrali. Oggi sappiamo che la spiegazione della struttura degli spettri non è connessa con campi magnetici interni all'atomo, ma con interazioni elettrostatiche che avvengono nell'atomo. Tuttavia l'idea di Ritz della precessione magnetica dell'elettrone prefigura in qualche misura il concetto di momento magnetico orbitale e di spin nella teoria quantistica dell'atomo.[16]

### VI—Il modello atomico di Bohr

Da buon sperimentatore qual era, quando nel 1911 introdusse il modello nucleare dell'atomo per spiegare i dati delle famose esperienze sulla diffusione delle particelle alpha, Rutherford non si preoccupò eccessivamente dei problemi insorgenti, che per i teorici costituivano un grave difetto del modello. Per esempio, non si poteva sfuggire al problema della stabilità, come era stato possibile a Thomson. Vi erano ormai prove convincenti che il numero ($n$) degli elettroni nell'atomo non poteva essere molto grande (doveva essere dell'ordine della metà del peso atomico $A$: $n \approx A/2$); almeno nel caso dell'idrogeno non era dunque evitabile la conclusione che l'unico elettrone, frenato nel suo moto dalla reazione di radiazione, avrebbe compiuto delle orbite che si restringevano sempre più, finendo per subire un collasso sul nucleo. La traiettoria percorsa dall'elettrone, partendo da una distanza dell'ordine di $10^{-8}$ cm (raggio dell'atomo) e fermandosi sul nucleo a circa $10^{-12}$ cm dal centro (raggio del nucleo), sarebbe apparsa come una spirale ad un osservatore esterno. A beneficio dei lettori interessati abbiamo riportato nell'*Appendice III* un calcolo del tempo di collasso dell'atomo di Rutherford.

Evidentemente la frequenza della radiazione emessa, venendo a dipendere dal raggio variabile dell'orbita a spirale percorsa dall'elettrone durante la caduta verso il nucleo, doveva variare con continuità, dando luogo ad uno spettro continuo. Gli spettri a righe rimanevano dunque assolutamente inspiegabili nel modello di Rutherford.

Quanto al problema dimensionale, veniva ad aggiungersi a quello dell'atomo anche il problema delle dimensioni del nucleo.

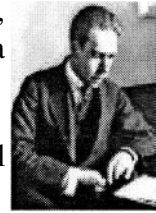

**N. Bohr
(1885-1962)**

Bohr, che lavorava nel laboratorio di Rutherford, ebbe la geniale intuizione di usare la costante $h$ di Planck (che ha le dimensioni di azione = energia x tempo) per costruire una grandezza atomica, dipendente soltanto dalla massa e dalla carica dell'elettrone, avente le dimensioni di una lunghezza:

$$\frac{(\text{azione})^2}{(\text{carica})^2 \text{ x massa}} = \text{lunghezza} .$$

Alcuni anni prima (1906) Jeans aveva fatto notare che combinando la massa e la carica dell'elettrone con l'energia cinetica si può ottenere il reciproco di una lunghezza: $\frac{\text{energia}}{(\text{carica})^2} = (\text{lunghezza})^{-1}$, ma aveva subito scartato la grandezza così costruita - che nelle sue intenzioni doveva determinare i numeri d'onda spettrali - perché era una funzione dell'energia, e questa può cambiare nel corso del tempo mentre i numeri d'onda spettrali osservati non cambiano.

Non appena Bohr vide la struttura "differenziale" della formula di Rydberg, com'egli stesso riferisce, tutto gli divenne chiaro: poteva usare la costante di Planck per ottenere una relazione energia-frequenza e, attraverso questa, pervenire a un'interpretazione teorica del principio di combinazione, in cui ogni termine corrisponde ad uno stato energetico dell'atomo.

---

[16] il concetto di spin dell'elettrone si può introdurre anche nell'elettrodinamica classica (v. M. K.-H. Kiessling, *Classical electron theory and conservation laws,* Phys. Lett. A258 (1999) 197-204).

Il modello atomico di Bohr (1913) è il primo modello quantistico dall'atomo, anche se le idee quantistiche vengono introdotte all'interno di un impianto ancora classico. Un approccio di questo tipo viene definito "*semiclassico*"*,* poiché il sistema in esame viene concepito come un oggetto classico (le posizioni e le velocità di tutte le particelle del sistema sono simultaneamente determinate istante per istante), e quindi viene descritto dalle leggi della meccanica di Newton e dell'elettrodinamica di Larmor-Lorentz; nel modello vengono poi introdotti "ad hoc" due postulati quantistici che permettono di spiegare la stabilità dell'atomo e gli spettri atomici.

Consideriamo il caso semplice di un atomo d'idrogeno (o uno ione idrogenoide) con l'unico elettrone che percorre orbite circolari che obbediscono alle leggi di Keplero, come in un piccolo sistema planetario. Come ora sapete, una descrizione classica interamente coerente del moto dell'elettrone prevede il collasso dell'elettrone sul nucleo lungo una traiettoria a spirale per effetto della reazione di radiazione. Ma mettiamo momentaneamente da parte questi effetti elettrodinamici e trascuriamo anche gli effetti relativistici (supponiamo cioè che la velocità dell'elettrone sia molto più piccola di quella della luce); inoltre assumiamo il nucleo puntiforme ed immobile. Con queste approssimazioni il moto dell'elettrone obbedisce alle semplici leggi di Newton.

Un semplice calcolo[17] mostra che la velocità dell'elettrone è data da

$$(10) \quad v = \sqrt{\frac{Ze^2}{mr}},$$

e la sua energia totale è

$$(11) \quad E = -\tfrac{1}{2}\frac{Ze^2}{r}.$$

La (11) afferma in pratica che il lavoro che si ricava arrestando l'elettrone (cioè $\tfrac{1}{2}\frac{Ze^2}{r}$) è minore del lavoro necessario per portare l'elettrone a distanza infinita dal nucleo vincendo la mutua attrazione elettrostatica (cioè $\frac{Ze^2}{r}$); in altre parole si deve compiere un lavoro pari $\tfrac{1}{2}\frac{Ze^2}{r}$ per ionizzare l'atomo. Afferma inoltre che l'atomo si trova in uno stato legato per qualunque valore finito di *r*; perciò se i fenomeni atomici obbedissero alla ordinaria meccanica newtoniana non vi sarebbe alcun limite superiore alle dimensioni atomiche, contro ogni evidenza sperimentale.

---

[17] muovendosi lungo la sua orbita circolare l'elettrone è soggetto a un'accelerazione $\frac{v^2}{r}$ diretta radialmente verso il nucleo ($v$ e $r$ sono rispettivamente la velocità e il raggio orbitale dell'elettrone). Da parte loro, l'elettrone e il nucleo interagiscono elettrostaticamente, di modo che agisce sull'elettrone una forza $\frac{Ze^2}{r^2}$ diretta dall'elettrone verso il nucleo (Z è il numero di protoni del nucleo, *e* è la carica elettrica del protone), e naturalmente una forza uguale e contraria agisce sul nucleo (ma ricordate che consideriamo immobile quest'ultimo). Per la seconda legge di Newton ( (forza centrifuga) = $m\frac{v^2}{r}$ = (forza coulombiana) = $\frac{Ze^2}{r^2}$ ) la velocità dell'elettrone è allora data da $v = \sqrt{\frac{Ze^2}{mr}}$, e la sua energia totale, uguale alla somma dell'energia cinetica $\tfrac{1}{2}mv^2 = \tfrac{1}{2}\frac{Ze^2}{r}$ e dell'energia potenziale elettrostatica $-\frac{Ze^2}{r}$ vale di conseguenza $E = -\tfrac{1}{2}\frac{Ze^2}{r}$.

Bohr risolse questo paradosso introducendo "ad hoc" la "*regola di quantizzazione*" del momento della quantità di moto (o momento angolare) dell'elettrone, secondo la quale il momento angolare dell'elettrone è un multiplo intero della costante universale $\frac{h}{2\pi}$:

$$(12)\ L = n\frac{h}{2\pi},$$

dove

$$(12')\ L = mvr, \text{ con } n=1,2,3,..., \quad h = 6.63 \cdot 10^{-27}\ \text{erg} \cdot \text{s.}^{18}$$

Questa regola limita il numero delle orbite possibili, tra quelle previste secondo la meccanica classica. Le orbite permesse dalla regola di quantizzazione (chiamate *"orbite quantiche"* o "*autoorbite*") sono infinite ma discrete (non continue, distanziate l'una dall'altra); a ciascuna di esse viene associato un numero intero *n*. *Le orbite quantiche sono stazionarie (corrispondono a stati stazionari dell'atomo), cioè l'elettrone può permanere indefinitamente su un'orbita quantica (senza perdere energia per radiazione elettromagnetica), con una stabilità che non può essere spiegata dalla meccanica classica.* Questo è il primo postulato di Bohr.

Il secondo postulato afferma, in totale contraddizione con l'elettrodinamica classica, che *l'irraggiamento può avvenire soltanto durante una transizione - o "salto quantico " - tra due orbite quantiche, e questa radiazione ha le stesse proprietà di quella che verrebbe emessa secondo la teoria classica da una carica che oscilla armonicamente con una frequenza costante.* Bohr non enunciò alcuna ipotesi sul meccanismo della transizione.

La regola di quantizzazione del momento angolare (12) risolve immediatamente il problema delle dimensioni atomiche. Infatti dalla (10) e dalla (12') si può esprimere il momento angolare come:

$$(12'')\ L = \sqrt{Ze^2 mr}.$$

Allora, da quest'ultima espressione e dalla regola di quantizzazione (12), si trova che il raggio $a_n$ dell'orbita quantica associata allo stato *n* è dato da:

$$(13)\ a_n = \frac{n^2 h^2}{4\pi^2 Z e^2 m}.$$

Per l'idrogeno (Z=l) nello stato fondamentale (*n*=1) si ottiene la quantità $a_1 = \frac{h^2}{4\pi^2 e^2 m} \simeq 0.529 \cdot 10^{-8}\text{cm}$ chiamata "raggio della prima orbita di Bohr dell'idrogeno" o semplicemente *"raggio di Bohr",* che costituisce una stima delle dimensioni dell'atomo nello stato fondamentale. Questo valore numerico è dello stesso ordine di grandezza di quello ottenuto dai risultati di esperienze indipendenti (misure di viscosità dei gas[19]).

---

[18] la condizione di quantizzazione (12), espressa da Bohr nella forma generale $\oint p\, dq = nh$ (dove *p* e *q* sono il momento e le coordinate dell'elettrone), fu introdotta per la prima volta da Planck nel 1911. Nella forma (12) è valida per le orbite circolari.

[19] v. per es. J.H. Hildebrand, *Introduzione alla teoria cinetica molecolare,* Progresso tecnico, Milano 1965.

Dalla (11) e dalla (13), si trova infine che i possibili valori dell'energia totale sono dati dall'espressione:

$$(14)\; E_n = -\frac{2\pi^2 Z^2 e^4 m}{n^2 h^2}.$$

I valori di $E_n$ rappresentano i livelli energetici dei sistemi idrogenoidi. Abbiamo così finalmente incontrato, nello studio dell'atomo d'idrogeno, il concetto di livelli discreti di energia (figura(9)). Questo concetto si è rivelato valido per tutti gli atomi e costituisce uno dei principali fondamentali della fisica quantistica, nonché la base della nostra raffigurazione della scala dell'energia.

Durante il salto quantico tra due orbite, se si fa l'ipotesi che l'energia totale venga conservata, l'atomo emette una certa quantità di radiazione elettromagnetica, che Bohr per semplicità considerò emessa in un singolo atto, omogenea, di frequenza $v$ e di energia $hv$. Per la conservazione dell'energia, la radiazione emessa dall'atomo ha un'energia uguale alla differenza $E_n - E_{n'}$ tra i due livelli $n$ (più alto, iniziale) e $n'$ (più basso, finale) coinvolti nella transizione; questa radiazione monocromatica determina la formazione di un'unica riga spettrale, che obbedisce alla fondamentale relazione

$$(15)\; v = \frac{E_n - E_{n'}}{h}$$

tra la frequenza emessa dall'atomo e le energie degli stati stazionari iniziale ($n$) e finale ($n'$); nello spettro comparirà la riga *nesima* appartenente alla serie di ordine $n'$ della specie atomica considerata.

Notiamo, di passaggio, che per descrivere il processo d'emissione Bohr fa ricorso alla cosiddetta "seconda teoria" della radiazione di Planck[20]. Il processo radiativo consiste dunque, per Bohr come per Planck, di singoli atti di emissione distinti, nei quali l'energia varia per quanti discreti. Una volta rilasciata nello spazio, però, questa radiazione si comporta in modo puramente classico, descritto dalla teoria ondulatoria, laddove secondo il punto di vista einsteiniano del 1905 la radiazione libera si componeva di unità distinte e indipendenti, dalla doppia natura ondulatoria e corpuscolare. Bohr, come vedete, manteneva sulla teoria della radiazione una posizione conservatrice, ancora molto legata alla fisica classica, mentre l'idea di Einstein era assolutamente rivoluzionaria. Circa venti anni dopo Bohr e Einstein furono protagonisti di un epico dibattito sui fondamenti della meccanica quantistica, ma questa volta a ruoli invertiti.

Il legame tra la relazione di Bohr energia-frequenza (15) e il principio di combinazione emerge immediatamente se i termini spettroscopici vengono interpretati alla luce dei livelli energetici. Nel caso dei sistemi idrogenoidi questa correlazione viene espressa (a meno di una costante indipendente da

---

[20] ci riferiamo alla teoria della radiazione che Planck sviluppò tra il 1910 e il 1912 e che è nota come "seconda teoria" di Planck. Nella "seconda teoria" planckiana i quanti intervengono soltanto durante i processi d'emissione ed assorbimento, mentre la radiazione libera conserva il carattere ondulatorio e continuo previsto dalla teoria classica. La rivoluzionaria idea einsteiniana del 1905 dei quanti di luce come costituenti indipendenti della radiazione libera veniva rifiutata dalla stragrande maggioranza dei fisici di allora (con la sola notevole eccezione di Stark). Ancora nel 1906. nel suo trattato sulla radiazione termica, Planck prescindeva dai concetti quantistici, che pure erano nati dalla sua fondamentale intuizione del quanto d'energia nel 1900 (v. M. Planck, *La teoria della radiazione termica,* Franco Angeli, Milano 1999). Su questa interessante problematica segnaliamo il nostro precedente articolo *La leggenda del quanto centenario* [http://ulisse.sissa.it/bUlb0504002.jsp] e l'intervento di John S. Rigden sul numero di Aprile 2005 di *Physics World* [http://physicsweb.Org/articles/world/18/4/2/l]. Nella fondamentale memoria del 1913 Bohr considerava il lavoro di Einstein come un mero contributo alla teoria di Planck, e ancora nel 1922 respingeva l'idea dei quanti di luce.

*n*) nel modo seguente: $\tau_n = -\frac{E_n}{hc}$. Allora, dalla (4) e dalla (14), si trova subito che il numero d'onda della riga emessa dall'atomo idrogenoide nella transizione da *n* a *n'* è dato da:

$$(16)\, \tilde{\nu} = \tau_{n'} - \tau_n = \frac{2\pi^2 Z^2 e^4 m}{h^3 c}\left(\frac{1}{n'^2} - \frac{1}{n^2}\right).$$

Se confrontiamo la (16) con la generalizzazione di Ritz della formula di Balmer (2) abbiamo la prova che la costante di Rydberg $R = \frac{2\pi^2 e^4 m}{h^3 c}$ è una costante universale, come da lungo tempo sospettavano gli spettroscopisti. Un'analisi più attenta mostra che in realtà il valore della costante di Rydberg dipende dalla massa *M* del nucleo.[21]

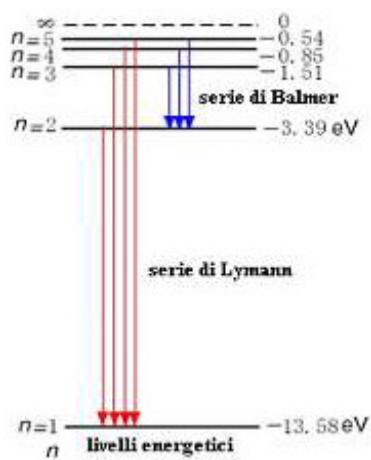

figura (9): livelli energetici dell'idrogeno

La teoria di Bohr è stata superata dai successivi sviluppi della meccanica quantistica[22] e della teoria dei campi. La prima, nelle sue diverse formulazioni - Meccanica delle matrici di Heisenberg, Born e Jordan (1925) e Meccanica Ondulatoria di Schrödinger (1926) - ha fornito una definizione concettualmente rigorosa degli stati di un sistema atomico. La seconda - in particolare l'Elettrodinamica Quantistica di Feynman, Schwinger e Tomonaga (1947) - è una raffinata teoria delle interazioni elettromagnetiche negli atomi. Ciononostante, la teoria "ingenua" di Bohr - nella versione generalizzata alle orbite ellittiche e includente gli effetti della relatività ristretta (Sommerfeld, 1916) - viene ancor oggi utilizzata dagli spettroscopisti perché rappresenta un buon compromesso tra semplicità e precisione.

---

[21] il valore teorico, calcolato nell'approssimazione della massa nucleare infinita (ossia nucleo immobile), è $R_\infty^{(\text{teor})} \simeq 109737\,\text{cm}^{-1}$. Il valore sperimentale per l'idrogeno, determinato spettroscopicamente (Paschen, 1916), è invece $R_H^{(\text{sperim})} \simeq 109678\,\text{cm}^{-1}$. Per tener conto del trascinamento del nucleo, cioè dell'effetto della sua massa finita, si deve applicare al valore teorico per la massa nucleare infinita una correzione, sostituendo alla massa *m* dell'elettrone la "massa ridotta" $\mu = \frac{mM}{m+M}$ del sistema elettrone-nucleo. Nel caso dell'idrogeno, dove $\mu \simeq 0.999456 \cdot m$, il valore teorico corretto $R_H^{(\text{teor})} = \frac{\mu}{M} R_\infty^{(\text{teor})} \simeq 109677\,\text{cm}^{-1}$ coincide pressoché perfettamente col valore sperimentale.

[22] per una recente rassegna di questi sviluppi suggeriamo la trilogia di A. Maccari, *La nascita della meccanica quantistica* (*I, II, III*), Giornale di Fisica, vol. 45, p. 129 (2004); vol. 45, p. 195 (2004); vol. 46, p. 81 (2005).

Il concetto dei livelli energetici resta del tutto valido e d'importanza preminente anche nella meccanica quantistica, dove il problema fondamentale rimane, come nella vecchia teoria di Bohr, la predizione/spiegazione dei livelli energetici dei sistemi microscopici (molecole, atomi, ioni atomici e molecolari, nuclei, particelle, quark, ...).

### *VII—Le collisioni anelastiche*

La prima evidenza diretta della scala atomica dell'energia risale al 1914, allorché i fisici J. Franck e G. Hertz condussero una ricerca sperimentale, che valse loro il Nobel per la fisica nel 1925, nella quale riuscirono a stimare le energie atomiche di eccitazione studiando l'urto con elettroni di atomi nello stato fondamentale. Per uno di quegli strani casi che talvolta accadono nella scienza, essi non erano consapevoli di aver in tal modo verificato la teoria di Bohr, che a quel tempo neppure conoscevano.[23] Per di più, la loro interpretazione dei risultati sperimentali non era del tutto corretta: essi credevano che le collisioni anelastiche tra gli elettroni e gli atomi di mercurio ionizzassero gli atomi, mentre in realtà tali collisioni provocano la transizione degli atomi di mercurio dallo stato fondamentale al primo stato eccitato.

Le collisioni elettrone-atomo possono essere elastiche o anelastiche. Nel primo caso la somma delle energie cinetiche dell'elettrone e dell'atomo si conserva, mentre nel secondo una parte dell'energia cinetica viene trasferita dall'elettrone all'atomo durante la collisione, e si trasforma in energia interna di eccitazione (se consideriamo l'elettrone puntiforme e privo di struttura interna la sua eccitazione non deve essere presa in considerazione). Poiché la massa di un atomo è almeno duemila volte la massa dell'elettrone, l'energia cinetica dell'atomo non cambia apprezzabilmente nella collisione, e si può considerare costante.

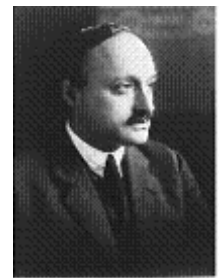

**J. Franck
(1882-1964)**

Di conseguenza, l'energia cinetica dell'elettrone e l'energia di eccitazione dell'atomo possono cambiare durante la collisione, ma non la loro somma. Se indichiamo con $K_i$ e $K_f$ le energie cinetiche dell'elettrone prima e dopo l'urto e con $W$ l'energia di eccitazione dell'atomo, allora:

a)  nel caso elastico abbiamo $K_i = K_f$ e $W = 0$,

b)  mentre nel caso anelastico $K_i > K_f$ e $W = K_i - K_f > 0$.

Indichiamo ora con $E_1, E_2, ...$ i livelli energetici dell'atomo, determinati spettroscopicamente, e supponiamo che prima dell'urto l'atomo si trovi nello stato fondamentale con energia $E_1$.

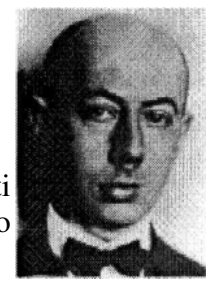

**G. Hertz
(1887-1975)**

Dopo l'urto l'atomo può trovarsi ancora nello stato fondamentale, oppure può andare in uno degli stati eccitati con energie $E_2, E_3, ...$ a seconda del valore dell'energia iniziale $K_i$ dell'elettrone urtante. Se $K_i < E_2 - E_1$ l'urto è elastico e l'atomo rimane nello stato fondamentale, se invece $K_i > E_2 - E_1$, l'urto è anelastico e l'elettrone urtante rimbalza via dall'atomo con energia $K_f = K_i + E_1 - E_2 \geq 0$, mentre l'energia dell'atomo aumenta di $W = E_2-E_1$. In particolare, quando $K_i=E_2-E_1$ l'energia dell'elettrone è appena sufficiente a portare l'atomo nel primo stato eccitato (chiamato anche livello di risonanza), e l'elettrone si arresta ($K_f =0$). Questa circostanza rende possibile una misura diretta dell'energia di risonanza. Franck ed Hertz, tabulando i valori di $K_f$ contro $K_i$, poterono determinare i valori di $E_2-E_1$ (cioè di $K_i$ quando $K_f = 0$).[24] In questo modo essi utilizzarono gli elettroni come sonde per investigare la struttura dell'atomo.

---

[23] v. G. Holton, *On the recent past of physics,* American Journal of Physics, vol. 29, p. 805 (1961).

Secondo la teoria quantistica, un atomo che è stato eccitato per urto elettronico nel livello di risonanza decade spontaneamente nello stato fondamentale emettendo un quanto di luce di frequenza $\nu = \dfrac{E_2 - E_1}{h}$. Il tempo di decadimento, che è dell'ordine di $10^{-8}$s, viene detto "vita media dello stato eccitato". Franck ed Hertz riuscirono anche a determinare simultaneamente la frequenza $\nu$ e l'energia del livello di risonanza $E_2$. Il loro apparato sperimentale era simile a quello utilizzato da Lenard negli studi sulla ionizzazione per urto, mostrato schematicamente nella figura (10).

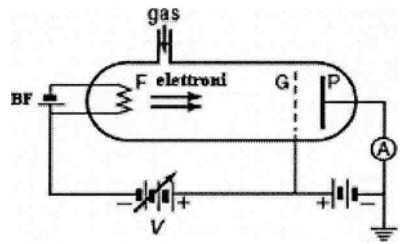

figura(10): apparato di Franck-Hertz

Del vapore di mercurio a bassa pressione viene introdotto in un tubo, al cui interno un filamento riscaldato F emette degli elettroni. Gli elettroni sono accelerati verso la griglia G, che si trova a un potenziale positivo rispetto a F; quelli che superano la griglia vengono rallentati da un piccolo potenziale ritardante, di circa 0.5-1 Volt, tra G e la placca P (la griglia si trova a un potenziale positivo rispetto alla placca) prima di raggiungere quest'ultima.

Gli elettroni, emessi per effetto termoionico con velocità trascurabile, vengono accelerati dalla differenza di potenziale (d.d.p.) F-G, dell'ordine di 10-20 Volt. La distanza G-P, di circa 1-2 millimetri, è molto minore del cammino libero medio degli elettroni nel gas contenuto nel tubo, mentre la distanza F-G è lievemente più grande del cammino libero medio[25]. Se indichiamo con $V$ la d.d.p. che accelera l'elettrone, questo raggiunge la griglia con energia $\tfrac{1}{2}\cdot mv^2 = eV$. Regolando $V$ si può fissare a piacimento l'energia degli elettroni che giungono in prossimità della griglia. Ammettendo che l'elettrone non perda energia lungo il percorso fino alla griglia, esso avrà un'energia $e(V-V_{GP})$ quando raggiunge la placca, dove il suo arrivo viene registrato come una variazione di corrente dal galvanometro A.

---

[24] il lavoro originale venne pubblicato su *Verhand. Deut. Physik Ges.*, vol. 16, pp. 457-467 (1914).
[25] queste condizioni sperimentali possono variare a seconda dei fenomeni che si vogliono mettere in evidenza negli esperimenti: urti elastici o anelastici, potenziali di eccitazione o di ionizzazione, ecc.. Inoltre vi sono vari fattori che influenzano i risultati degli esperimenti, come il numero e l'effettiva distribuzione delle velocità degli elettroni emessi dal catodo, l'accumulo di carica spaziale in prossimità del catodo. le distanze tra gli elettrodi, la geometria del tubo, i meccanismi di eccitazione, la formazione di ioni... (v. G. F. Hanne, *What really happens in the Franck-Hertz experiment with mercury?*, American Journal of Physics, vol. 56. p. 696 (1988)).

Poiché in queste esperienze si lavora con potenziali dell'ordine di alcuni volt, è preferibile per comodità esprimere l'energia degli elettroni in eV (l eV $\simeq 1.60 \cdot 10^{-19}$J). Come sappiamo dalla teoria quantistica, tra energia di eccitazione e lunghezza d'onda della radiazione emessa dall'atomo vale la relazione $E = eV = h\dfrac{c}{\lambda}$, da cui

$$(17) \quad \lambda V = \frac{hc}{e} \simeq 12395 \, ,$$

dove $\lambda$ è espresso in Å e $V$ in Volt. Mediante questa relazione i potenziali atomici di eccitazione si possono esprimere in termini di righe d'emissione o viceversa.

Quando il potenziale $V$ aumenta da 0 fino al valore $\dfrac{E_2 - E_1}{e}$ la corrente registrata dal galvanometro aumenta; in questo intervallo gli urti sono elastici e gli elettroni accelerati raggiungono la griglia con energia sufficiente a risalire con facilità il debole controcampo G-P. Ma non appena $V$ supera $\dfrac{E_2 - E_1}{e}$ gli elettroni cedono quasi tutta l'energia in urti anelastici con atomi in prossimità della griglia, e solo pochi elettroni riescono a risalire il controcampo. Il potenziale critico $V^I = \dfrac{E_2 - E_1}{e}$ viene detto *"potenziale di risonanza"*. Al di sopra di questo valore il punto nel quale gli elettroni raggiungono la velocità critica si sposta verso F e un numero sempre maggiore di elettroni subisce urti anelastici; la corrente registrata dal galvanometro diminuisce allora rapidamente, fino a un valore minimo quando $V$ raggiunge il valore $V^I + V_{GP}$.

Aumentando ancora $V$ da $V^I + V_{GP}$ a $2\,V^I$ la corrente riprende ad aumentare, perché ora gli elettroni in media sono soggetti a urti anelastici prima di raggiungere la griglia e hanno così il tempo di accelerare una seconda volta e risalire il debole controcampo G-P.

Quando $V$ raggiunge il valore $2\,V^I$, cioè il doppio del potenziale di risonanza, la corrente cala di nuovo velocemente. Infatti ora gli urti anelastici avvengono in media a metà strada tra F e G e gli elettroni hanno il tempo di riacquistare interamente l'energia perduta e subiscono un secondo urto anelastico prima di raggiungere la griglia. Si raggiunge un nuovo minimo della corrente quando $V$ è uguale a $2\,V^I + V_{GP}$; al di sopra di questo valore la corrente riprende ad aumentare.

Lo stesso comportamento della corrente si ripete per valori del potenziale accelerante pari a $3\,V^I$, $4\,V^I$, ecc. Come mostrato nella figura (11) i massimi sono tutti situati alla medesima distanza, data dal potenziale di risonanza $V^I$ (4.9V per il mercurio).

Per misurare *potenziali di eccitazione* d'ordine più elevato (del potenziale di risonanza) è necessario ridurre la pressione in modo che il cammino libero medio degli elettroni nel gas sia dell'ordine della distanza F-G. Un esempio delle caratteristiche corrente-d.d.p. che si possono ottenere è mostrato nella figura (11').

Quando la pressione del gas nel tubo viene fatta diminuire compaiono nuovi picchi nelle caratteristiche corrente-d.d.p. (prima $V^I$ =4.9V, poi $V^{II}$ =6.7V, poi $V^{III}$ =8,8V, ...). La pressione dev'essere regolata accuratamente; da questo dipende infatti il cammino libero medio degli elettroni nel gas e quindi la probabilità di collisione con gli atomi di mercurio. Adoperando un vapore saturo (ottenibile per esempio ponendo nel tubo una goccia di mercurio) la pressione dipende soltanto dalla temperatura. Per questo motivo durante gli esperimenti il tubo viene messo in un bagno di calore (un forno o un refrigeratore a seconda che interessino alte o basse pressioni) a temperatura costante, controllata mediante una termocoppia[26]; in tal modo la pressione del vapore saturo può essere completamente determinata e controllata.

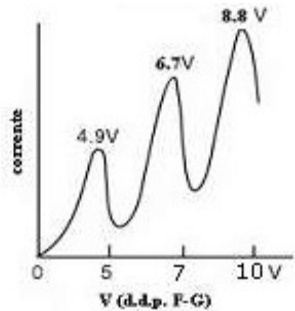 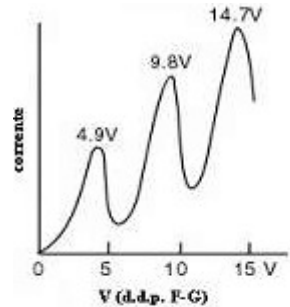

figura (11): caratteristica corrente-ddp    figura (11´): caratteristica corrente-ddp a bassa pressione

Alle pressioni estremamente basse il cammino libero medio e il più elevato potenziale di eccitazione che si può determinare con questo metodo sono fortemente influenzati dalla geometria del tubo.

Il potenziale di ionizzazione, cioè il lavoro (in eV) necessario per separare un elettrone dal resto dell'atomo (ione $Hg^+$), può essere misurato con la stessa apparecchiatura. A tale scopo viene stabilita tra G e P una d.d.p. molto elevata (circa 40 volt). In queste condizioni nessun elettrone può raggiungere P risalendo il controcampo G-P. Tuttavia, se qualche atomo viene ionizzato per urto elettronico in prossimità della griglia, lo ione $Hg^+$ residuo viene accelerato dal potenziale G-P e il galvanometro A registra una corrente. Questo accade per un determinato valore $V_i$ del potenziale acceleratore, chiamato *"potenziale di ionizzazione"* (per il mercurio $V_i$ =10.4V). Quando il potenziale acceleratore $V$ supera $V_i$, gli elettroni subiscono un urto anelastico e rimbalzano via con l'energia $K_f = V - V_i$, che può assumere tutti i valori possibili (spettro continuo).

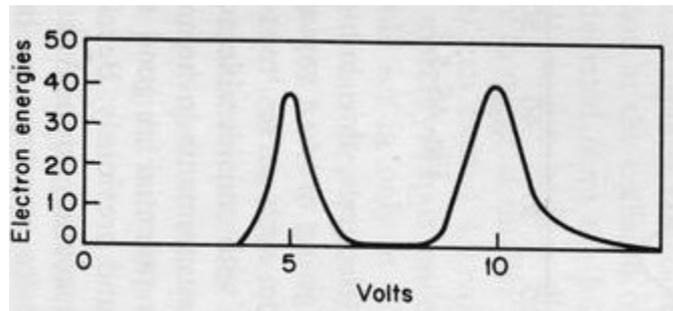

Nel resoconto dei loro esperimenti Franck ed Hertz presentarono vari diagrammi che mostrano, come nell'esempio qui a fianco, massimi molto evidenti in corrispondenza dei potenziali di eccitazione.

Questo ci suggerisce, anche se le cose sono un po' più complicate, di rappresentare approssimativamente l'andamento di $K_f$ contro $K_i$ con una funzione a denti di sega. Di conseguenza, la perdita di energia $K_i - K_f$ in funzione di $K_i$ avrà un profilo a scalini, come mostra la figura (12), che rappresenta visivamente la nostra scala dell'energia. Ve ne renderete facilmente conto osservando i due diagrammi qui a fianco: il primo rappresenta una funzione a denti di sega $y(x)$; il secondo, che mostra una funzione a scalini, si può costruire dal primo riportando i valori di $x-y$ contro $x$.

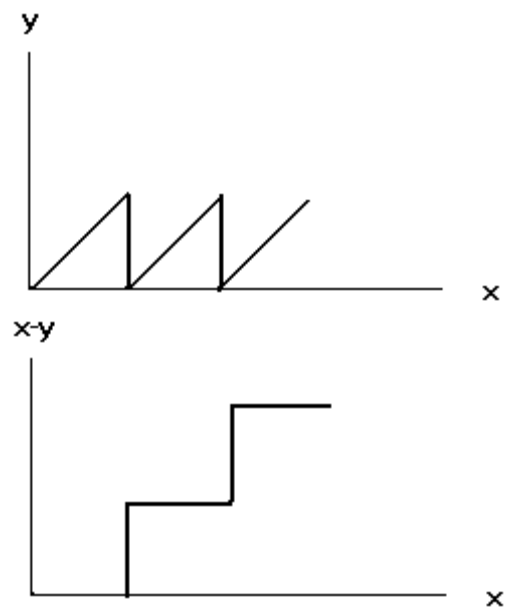

---

[26] i valori resistenza-temperatura per la termocoppia sono riportati in appositi *handbook*, ma al giorno d'oggi questi dati si possono trovare in rete [http:srdata.nist.gov/its90/main].

## VIII— *I livelli energetici*

In un'altra esperienza - sempre descritta nello stesso lavoro del 1914 - Franck ed Hertz proiettarono sulla fenditura di uno spettroscopio l'immagine della regione del tubo dove avvengono le collisioni elettrone-atomo. Regolando con un potenziometro il valore del potenziale accelerante V, essi poterono osservare la riga di risonanza del mercurio (allo stato di vapore a bassa pressione) $\lambda$=2536.6Å, che si trova nell'ultravioletto. Questa riga è dovuta al decadimento di atomi di mercurio eccitati nelle collisioni con elettroni. Dalla (17) si ottiene un valore teorico del potenziale di eccitazione pari a 4.89 Volt, contro il valore sperimentale di 4.9 Volt determinato dalle esperienze di collisione, in accordo con la teoria entro gli errori sperimentali. Questa determinazione del potenziale atomico di prima eccitazione venne eseguita fotograficamente nella regione UV.

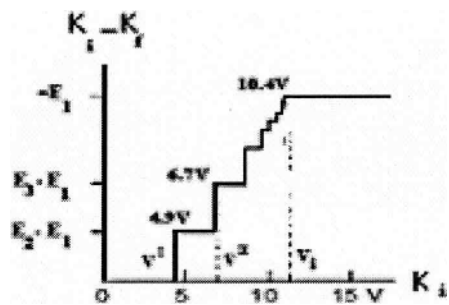

figura (12): perdita di energia negli urti anelastici

Successivamente sono stati adoperati metodi spettroscopici e fotoelettronici. Questi ultimi hanno confermato i risultati di Franck-Hertz, mettendo in evidenza un massimo di corrente fotoelettronica per il potenziale di eccitazione previsto. Queste, e numerose altre, conferme dei risultati pionieristici di Franck-Hertz hanno fornito solidi fondamenti sperimentali al concetto di livelli energetici atomici. Questo concetto ha permesso d'interpretare numerosi ed importanti fenomeni fisici per i quali la fisica classica si era dimostrata inadeguata a fornire una spiegazione. Diamo solo due esempi.

1. L'indagine spettroscopica su vapori di mercurio saturi a bassissima pressione, irradiati con una lampada a pareti di quarzo che emette uno spettro contenente la riga $\lambda$=2536.6 Å dell'UV, dimostra che gli atomi del mercurio non assorbono (cioè non convertono in calore) questa riga, ma la diffondono in tutte le direzioni. La luce diffusa è molto intensa e altamente monocromatica. Questo fenomeno, detto "risonanza ottica", venne scoperto dal celebre fisico americano R.W. Wood (1902).

      1c.        Spiegazione classica.

Il vapore di mercurio contiene degli oscillatori, alcuni dei quali hanno una lunghezza propria $\lambda$=2536.6 Å, che quando vengono investiti da luce con la lunghezza d'onda appropriata risuonano meccanicamente ed irradiano in tutte le direzioni un'intensa luce di lunghezza d'onda uguale alla lunghezza propria degli oscillatori.

      1q.        Spiegazione quantistica.

La luce di lunghezza d'onda $\lambda$=2536.6 Å induce transizioni di atomi di mercurio nel livello di risonanza; questi atomi decadono spontaneamente (se il tempo tra due successive collisioni con elettroni è lungo rispetto alla vita media del livello di risonanza) dopo circa $10^{-8}$s, e viene emessa in tutte le direzioni luce della riga $\lambda$=2536.6 Å. Lo stesso fenomeno si manifesta per i vapori di sodio irradiati con luce gialla del doppietto D ($\lambda$=5890/6 Å).

2. Certe sostanze, irradiate con luce monocromatica, riemettono luce di natura diversa in tutte le direzioni. Questa luce però non contiene frequenze più alte di quelle della luce eccitatrice (legge empirica di Stokes (1852)). Questo fenomeno, noto come "fluorescenza", ha un'immediata interpretazione quantistica:

    2q.    Spiegazione quantistica.

La luce eccitatrice porta l'atomo in uno stato eccitato. Se il decadimento nello stato fondamentale consiste in una cascata di salti quantici, in ogni salto viene emesso un fotone; questo (poiché nella cascata l'energia si conserva) non può avere un'energia (e quindi una frequenza) maggiore di quella del fotone eccitatore.

*Appendici*

*I- Tempo di decadimento dell'atomo "classico"*

Secondo l'elettrodinamica classica, la potenza irradiata da una carica puntiforme, come un elettrone, in moto accelerato (se consideriamo solo il contributo della radiazione di dipolo elettrico e trascuriamo gli effetti relativistici) obbedisce alla legge di Larmor

$$(A)\ P = \frac{2}{3} \cdot \frac{e^2 a^2}{c^3}$$, dove $a$ è l'accelerazione della carica[27] (Larmor, 1897).

Indichiamo con $W_0$ l'energia totale (cinetica + potenziale) iniziale dell'elettrone prima che venga accelerato e cominci ad irradiare. Il processo di radiazione farà diminuire l'energia dell'elettrone ad un valore $W < W_0$, mentre, per la conservazione dell'energia, la differenza $W_0$-$W$ si troverà dispersa nello spazio sotto forma di radiazione. In ogni istante l'energia dell'elettrone è uguale alla somma $W=½·mv^2 +U$ dell'energia cinetica ($½·mv^2$) e dell'energia potenziale $(U)$ immagazzinata nel campo. La conservazione dell'energia impone che $W+(W_0-W) =½·mv^2 + U + (W_0 -W) = W_0$ =cost. Ne risulta, dall'annullarsi della derivata temporale, la relazione $m\mathbf{a} \cdot \mathbf{v} - \mathbf{F} \cdot \mathbf{v} + P = (m\mathbf{a} - \mathbf{F}) \cdot \mathbf{v} + P = 0$[28], che lega $m$, $\mathbf{a}$, $\mathbf{v}$ (rispettivamente massa, accelerazione e velocità dell'elettrone), la forza $\mathbf{F}$ che agisce sulla carica (dovuta al potenziale in cui si muove l'elettrone), e la potenza emessa $P$. Si può interpretare $m\mathbf{a}$-$\mathbf{F}$ come una forza di reazione $\mathbf{F_r}$ sulla carica, che deve essere presente (dato che l'energia totale è conservata) per spiegare la perdita $P$ per irraggiamento. Ora. la forza totale $\mathbf{F+F_r}$ è legata all'accelerazione a dalla seconda legge di Newton $\mathbf{F+Fr}=m\mathbf{a}$. Possiamo allora scrivere $\mathbf{Fr} \cdot \mathbf{v}$ = -P $= -\frac{2}{3} \cdot \frac{e^2 a^2}{c^3}$, dove il segno negativo indica che la forza di reazione si oppone al moto della carica. Si può dimostrare che, sotto certe condizioni,[29] $\mathbf{F_r}$ si può esprimere come $(B) \mathbf{F_r} = \frac{2}{3}\frac{e^2}{c^3}\mathbf{\dot{a}}$, dove $\mathbf{\dot{a}}$ è la derivata temporale dell'accelerazione. Si noti che se l'accelerazione è uniforme ($\mathbf{\dot{a}} = 0$) la forza di reazione si annulla e la carica non dovrebbe irradiare.[30]

---

[27] v. p.es. la voce *Radiazione elettromagnetica (teoria della),* di F. Tito Arecchi, nell' *Enciclopedia della scienza e della tecnica,* Vol. X, Mondadori, Milano 1980[(7)].

[28] $\frac{d}{dt}(½ mv^2) = m\mathbf{v} \cdot \frac{d\mathbf{v}}{dt} = m\mathbf{a} \cdot \mathbf{v},\ \frac{dU}{dt} = \frac{dU}{d\mathbf{r}} \cdot \frac{d\mathbf{r}}{dt} = -\mathbf{F} \cdot \mathbf{v},\ \frac{d(W_0 - W)}{dt} = -\frac{dW}{dt} = P$.

[29] v. W.H.K. Panofsky- M. Phillips, *Elettricità e Magnetismo,* Ambrosiana, Milano 1966.

La (B) ha due conseguenze importanti per la stabilità dell'atomo classico.

1. Un elettrone legato ad un atomo con un potenziale armonico oscilla con un moto armonico, smorzato dalla reazione di radiazione $\mathbf{F_r}$. Se consideriamo per semplicità il moto dell'elettrone in una sola dimensione, la forza elastica di legame è data da $F = -kl$, dove $l$ è lo spostamento dell'elettrone dalla posizione di equilibrio. Allora, per la seconda legge di Newton abbiamo $ma + kl = F_r = \frac{2}{3}\frac{e^2}{c^3}\dot{a}$ e dividendo per $m$ giungiamo all'equazione del moto $a + \omega_0^2 l = \frac{2}{3}\frac{e^2}{mc^3}\dot{a} = \frac{\gamma}{\omega_0^2}\dot{a}$, dove $\gamma = \frac{2}{3}\frac{e^2\omega_0^2}{mc^3}$ e $\omega_0^2 = \frac{k}{m} = 2\pi\nu_0$ ($\nu_0$ = frequenza naturale dell'oscillatore). Se la forza di reazione è molto piccola rispetto alla forza elastica che fa oscillare la carica[31], l'ampiezza dell'oscillazione diminuirà secondo la legge, valida per $\gamma$ piccolo $l(t) \simeq l_0 e^{-i\omega_0 t - \gamma t/2}$, dove $l_0$ è l'ampiezza massima. Troviamo allora che l'energia totale $W$ dell'oscillatore diminuisce secondo la legge $W = W_0 e^{-\gamma t} = \frac{1}{2} m\omega_0^2 l_0^2 e^{-\gamma t}$, e la potenza irradiata è data da

$$(C) \; P = -\frac{dW}{dt} = \gamma W.$$

Dalla (C) segue immediatamente $\frac{1}{\gamma} = \frac{W}{P}$, che rappresenta la durata media dell'impulso irradiato, cioè il tempo impiegato per dissipare l'energia dell'oscillatore. Questa relazione dice anche che la potenza istantanea irradiata è proporzionale all'energia posseduta dall'oscillatore in quel momento.

Vediamo quindi che l'ampiezza di oscillazione diminuisce nel tempo come $l(t) \propto e^{-\gamma t/2}$, con un tempo medio di smorzamento dato da $\frac{1}{\gamma}$.

2. Analogamente, la potenza irradiata $P = \gamma W = \frac{2}{3}\frac{e^2\omega_0^2}{mc^3} \cdot \frac{1}{2} m\omega_0^2 l_0^2 e^{-\gamma t} = \frac{e^2\omega_0^4}{3c^3} l_0^2 e^{-\gamma t}$ aumenta con la quarta potenza della frequenza (e diminuisce nel tempo) come $P \propto \omega_0^4 e^{-\gamma t}$.

Su queste basi possiamo esprimere il tempo medio di decadimento di un dipolo elettrico atomico come $\tau_0 = \frac{1}{\gamma} = \frac{3}{2}\frac{mc^3}{e^2} \cdot \frac{1}{\omega_0^2} = \frac{3}{2}\frac{c}{r_0} \cdot \frac{1}{\omega_0^2} \simeq 10^{23} \text{s}^{-1} \cdot \frac{1}{\omega_0^2}$, dove $r_0 = \frac{e^2}{mc^2} \simeq 2.82 \cdot 10^{-13}$cm è il raggio classico dell'elettrone.

I risultati precedenti sono validi per $\gamma \ll \omega_0$, ovvero $\gamma \ll \omega_0 \ll \frac{c}{r_0} \sim 10^{23}$Hz.[32] Questa frequenza limite $\sim 10^{23}$Hz è estremamente alta rispetto alle frequenze ottiche (per confronto, la frequenza della riga H$_\alpha$ dell'idrogeno è $\nu_0 \simeq 0.5 \cdot 10^{15}$ Hz, ovvero $\omega_0 \simeq 3.14 \cdot 10^{15}$ Hz, che è inferiore di otto ordini di grandezza alla frequenza limite) e il suo reciproco è un tempo limite $\sim 10^{-23}$s che corrisponde al decadimento più veloce possibile (entro i limiti di validità del calcolo). In termini relativi, pertanto, lo smorzamento delle oscillazioni alle frequenze ottiche avviene molto lentamente (oscillazioni quasi-elastiche). Il calcolo dimostra che alle frequenze ottiche il decadimento dura in media $\tau_0 \sim 10^{-8}$s. Questo può sembrare un tempo molto breve, ma in termini relativi è lunghissimo (ben $10^{15}$ volte più lungo del tempo limite sopra definito $\sim 10^{-23}$s).

---

[30] v. K.H. Mariwalla - N.B. Hari Dass, *A Hundred Years of Larmor Formula,* Preprint No: IMSc/98/01/03, Physics Education (http:ArXiv.org/physics/0205046).

[31] se la forza di reazione è molto piccola - quindi se $\frac{\gamma}{\omega_0^2}$ è molto piccolo - l'equazione del moto $a + \omega_0^2 l - \frac{\gamma}{\omega_0^2}\dot{a} = 0$ può essere approssimata con $a + \omega_0^2 l \simeq 0$. Derivando quest'ultima rispetto al tempo otteniamo, per la derivata temporale dell'accelerazione $\dot{a} \simeq -\omega_0^2 v$, che inserita nell'equazione di partenza porta alla nuova equazione approssimata $a + \gamma v + \omega_0^2 l = 0$, con soluzione $l(t) \simeq l_0 e^{-i\omega_0 t - \gamma t/2}$.

[32] come si ricava immediatamente da $\gamma \sim \frac{r_0}{c}\omega_0^2 \ll \omega_0$.

Aggiungiamo che ogni riga presenta un caratteristico allargamento, che corrisponde a $\gamma$ e vale, alle frequenze ottiche, $\Delta\omega \simeq \gamma \sim 10^8$ Hz se espresso in termini di frequenza, e $\Delta\lambda = \dfrac{2\pi\Delta\omega c}{\omega_0^2} \simeq \dfrac{2\pi\gamma c}{\omega_0^2} = \dfrac{4\pi}{3} r_0 \sim 10^{-12}$cm, indipendente dalla frequenza, in termini di lunghezza d'onda.

## II — Interpretazione classica dell'effetto Zeeman

L'interpretazione classica dell'effetto Zeeman normale si fonda sulla teoria del moto in un campo magnetico di una particella carica legata elasticamente o quasi-elasticamente. In presenza di un campo magnetico esterno **H** sull'elettrone agisce la forza di Lorentz $-\dfrac{e}{c}\mathbf{v}\times\mathbf{H}$ e il moto dell'elettrone (assumendo che avvenga in un piano perpendicolare al vettore **H**) viene descritto dal sistema di equazioni differenziali $(A)\begin{cases}\ddot{x}+\omega_0^2 x+2\omega_L\dot{y}=(\gamma/\omega_0^2)\dddot{x}\\ \ddot{y}+\omega_0^2 y+2\omega_L\dot{x}=(\gamma/\omega_0^2)\dddot{y}\end{cases}$, dove $\omega_L=\dfrac{eH}{2mc}$ corrisponde alla frequenza di Larmor $\nu_L=\dfrac{\omega_L}{2\pi}$ e $\gamma\ll\omega_0$. Per queste equazioni si possono cercare soluzioni della forma $\begin{cases}x=x_0 e^{-i\omega t-\gamma t/2}\\ y=y_0 e^{-i\omega t-\gamma t/2}\end{cases}$, che introdotte nel sistema (A) lo trasformano, dopo le approssimazioni del caso, nel sistema di equazioni lineari algebriche $\begin{cases}(\omega_0^2-\omega^2)x_0-2i\omega_L\omega y_0=0\\ 2i\omega_L\omega x_0+(\omega_0^2-\omega^2)y_0=0\end{cases}$. Questo può essere risolto quando il determinante dei coefficienti si annulla, ovvero quando $(\omega_0^2-\omega^2)^2-4\omega_L^2\omega^2=0$. Si ottiene in tal modo un'equazione di secondo grado in $\omega^2$, che ha due soluzioni: $\omega_{1,2}^2=\omega_0^2+2\omega_L^2\pm 2\omega_L\sqrt{\omega_0^2+\omega_L^2}$. Per un campo debole ($\omega_L\ll\omega_0$), in prima approssimazione risulta $\omega_{1,2}\simeq\omega_0\pm\omega_L$, mentre quando il campo è intenso ($\omega_L>\omega_0$), si ricava per le due righe separate, sempre dopo le approssimazioni del caso[33], $\omega_1\approx 2\omega_L,\quad \omega_2\approx\dfrac{\omega_0^2}{2\omega_L}$.

## III- Tempo di collasso dell'atomo di Rutherford

Consideriamo nuovamente la perdita d'energia dell'elettrone per radiazione (legge di Larmor)

$$(A)\ P=\dfrac{2}{3}\cdot\dfrac{e^2 a^2}{c^3}.$$

Il moto dell'elettrone intorno al nucleo è influenzato dalla reazione di radiazione che abbiamo già considerato nel caso quasi-elastico; supponiamo che anche nel caso presente la forza di reazione sia molto piccola e che il nucleo, molto più pesante dell'elettrone, possa essere considerato immobile; supponiamo infine, nell'approssimazione adiabatica, che l'orbita dell'elettrone intorno al nucleo si restringa pochissimo, rimanendo quasi circolare in ogni istante. Con queste approssimazioni possiamo ricavare l'accelerazione dell'elettrone dall'equazione newtoniana del moto $ma = \dfrac{e^2}{r^2}$, e sostituirla nella (A) per ottenere la perdita d'energia nell'unità di tempo

$$P = \frac{2}{3} \cdot \frac{e^6}{m^2 c^3 r^4}.$$

---

[33] si applica lo sviluppo in serie: $\sqrt{1+x}=1+\tfrac{1}{2}x-\tfrac{1}{8}x^2+\ldots$ per $x<1$.

D'altra parte nel caso considerato (non-relativistico) l'energia totale dell'elettrone è data da $W = -\frac{1}{2}\dfrac{e^2}{r}$ da cui si ricava $\dfrac{dW}{dr} = \frac{1}{2}\dfrac{e^2}{r^2}$. In conseguenza della perdita d'energia l'elettrone rallenta e il raggio dell'orbita decresce in funzione del tempo. Allora $P = -\dfrac{dW}{dt} = -\dfrac{dW}{dr} \cdot \dfrac{dr}{dt} = -v \cdot \dfrac{dW}{dr}$ da cui $v = \dfrac{dr}{dt} = -P/(dW/dr) = -\frac{4}{3}\dfrac{e^4}{m^2 c^3 r^2}$. Integrando quest'ultima equazione su $r$ da $R \simeq 0.5 \cdot 10^{-8}$ cm (raggio dell'atomo d'idrogeno) a $r \simeq 0$ (poiché il raggio del nucleo è inferiore al raggio atomico di quattro ordini di grandezza possiamo considerarlo nullo in prima approssimazione) e integrando su $t$ da 0 a $\tau$, otteniamo un tempo di collasso

$$\tau = \frac{1}{4}\frac{m^2 c^3}{e^4} R^3 \simeq 1.3 \cdot 10^{-11}\,\text{s}.$$

Questo significa che nell'atomo di Rutherford il collasso dell'elettrone avviene con estrema rapidità, circa 1000 volte più velocemente che nel caso quasi-elastico. Si tratta di un tempo dello stesso ordine di grandezza della vita media di uno stato eccitato dell'idrogeno, il cui stato fondamentale sembra invece avere una vita media infinita.

## *Bibliografia*[34]

---

[34] nella stesura della prima parte di questo articolo ci siamo avvalsi parzialmente dell'ipertesto di A. Garuccio [http:www.ba.infn.it/~garuccio/didattica/spettroscopia/indice.html], che presenta un'introduzione didattica alla storia della spettroscopia. Un'altra utile risorsa in rete sono le *Nobel Lectures* di Bohr, Franck ed Hertz [http:nobelprize.org/physics/laureates].